\documentclass[aps,prd,superscriptaddress,nofootinbib,preprintnumbers]{revtex4-1}
\usepackage{amsmath}
\usepackage{graphicx}
\usepackage{subfigure}
\usepackage{amssymb}
\usepackage{xcolor}
\usepackage{multirow}
\usepackage{cancel}
\usepackage{color}
\usepackage{ulem}
\usepackage{caption}
\usepackage{listings}
\usepackage{xcolor}
\usepackage{float}
\usepackage{slashed}

\usepackage{graphicx}

\usepackage{float}
\usepackage{amsmath}
\usepackage{amssymb}
\usepackage[utf8]{inputenc}
\usepackage{hyperref}
\usepackage{color}
\usepackage{slashed}
\allowdisplaybreaks

\newcommand{\be}{\begin{equation}}
\newcommand{\ee}{\end{equation}}

\begin{document}

\title{Automatic detection of boosted Higgs boson and top quark jets in an event image}

\author{Sang Kwan Choi}
\email{hermit1231@sogang.ac.kr}
\affiliation{College of Physics, Sichuan University, Chengdu 610065, China}

\author{Jinmian Li}
\email{jmli@scu.edu.cn (corresponding author)}
\affiliation{College of Physics, Sichuan University, Chengdu 610065, China}

\author{Cong Zhang}
\email{zhangcong.phy@gmail.com}
\affiliation{College of Physics, Sichuan University, Chengdu 610065, China}

\author{Rao Zhang}
\email{zhangrao@stu.scu.edu.cn}
\affiliation{College of Physics, Sichuan University, Chengdu 610065, China}

\begin{abstract}
We build a deep neural network based on the Mask R-CNN framework to detect the Higgs jets and top quark jets in any event image. 
We propose an algorithm to assign the top quark final states at the ground truth level so that the network can be trained in a supervised manner.
A new jet branch is added to the network, which uses constituent information to predict the four-momenta of the original parton, thus intrinsically implementing the pileup mitigation.
The network can predict both the shapes and the momenta of target jets.
We show that the network surpasses the LorentzNet in top and Higgs tagging and the PELICAN network in momentum regression for certain cases, in terms of reconstruction efficiency and accuracy.
We also show that the performance of the network does not degrade much when applied to events from a different process from the trained one and to events with overlapping jets. 
\end{abstract}


\maketitle

\newpage
\section{Introduction}\label{sec:intro} 

Due to the color confinement, the quarks and gluons produced in a hard process cannot be detected individually. 
Instead, it will go through parton showering and hadronization after production, giving rise to a collimated spray of energetic detectable color singlet hadrons, which is referred to as a jet. 
Establishing the correspondence between jets and partons is essential for understanding the underlying physics of collider events. 
It requires an infrared-safe algorithm that can attribute the related final state hadrons to its partonic ancestors and predict the four-momentum of the parton. 

At the LHC and other collider experiments, the jets are reconstructed by using sequential recombination algorithms~\cite{Salam:2010nqg}, in which the final state hadrons are pairwise recombined according to some distance measures, such as Cambridge/Aachen algorithm~\cite{Dokshitzer:1997in}, anti-$k_{T}$ algorithm~\cite{Cacciari:2008gp} and so on~\cite{Czakon:2022wam,Gauld:2022lem,Caola:2023wpj}. 
In those jet clustering algorithms, an appropriate cone size parameter needs to be taken according to the configuration of the detector and the properties of the target jet. 
At the LHC, the anti-$k_{T}$ jet algorithm with cone size $R=0.4$ works efficiently in finding quark and gluon jets in the ATLAS and CMS detectors. 
Another origin of a jet is a boosted heavy particle decaying into hadronic final states, for example, the top quark and the Higgs boson. 
As the typical jet cone size of a heavy resonance is given by $R \sim 2 m/p_{T}$, a much larger value is adopted to fully capture the jet constituents. 
In contrast to the jet originating from the light quark and gluon, those fat-jets are characterized by remarkable substructure. A variety of jet substructure techniques~\cite{Abdesselam:2010pt,Altheimer:2012mn,Altheimer:2013yza,Adams:2015hiv,Larkoski:2017jix,Kogler:2018hem} have been proposed for tagging heavy resonant jets, such as mass-drop tagging~\cite{Butterworth:2008iy} for the Higgs boson, HEPTopTagger algorithm~\cite{Plehn:2010st} for the top quark, and N-subjettiness~\cite{Thaler:2010tr,Thaler:2011gf} for general fat-jets. 
One practical issue for the fat-jet reconstruction at hadron colliders would be the heavy contamination from pileup events as well as underlying events. 
The average number of pileup events reaches $\langle \mu \rangle=35$ for LHC Run-II and $\langle \mu \rangle=140$ for High-Luminosity LHC. 
The large cone size of fat-jet renders many of those background particles to be included as the jet constituents. 
Although the jet grooming methods~\cite{Krohn:2009th,Ellis:2009su} have been found to be very helpful in mitigating the pileup effects, there are still relatively large errors in obtaining the original parton momentum when it is calculated by the vector sum of momenta of the groomed jet constituents. 

In terms of fat-jet tagging efficiency, machine-learning techniques have proven to substantially outperform those jet substructure techniques~\cite{Larkoski:2017jix,Guest:2018yhq,Albertsson:2018maf,Radovic:2018dip,Feickert:2021ajf}.
According to the jet formation, a jet can be either viewed as sequences/trees formed through sequential parton showering and hadronization or viewed as graphs/point clouds with the information encoded in the adjacency nodes and edges. Moreover, the calorimeters inside the detector measure the angular position and energy of particles on fine-grained spatial cells. Considering each calorimeter cell as a pixel and the energy deposition as the intensity, a jet can be naturally viewed as a digital image. 
All of those three representations of the jet are common objects in machine learning. They can be proceeded by different kinds of neural network architectures, {\textit i.e.} recurrent neural networks~\cite{Andreassen:2018apy} and transformer network~\cite{vaswani2017attention} for sequence, graph neural networks (GNN)~\cite{Moreno:2019bmu,Komiske:2018cqr,Qu:2019gqs} for point cloud, recursive neural networks~\cite{Louppe:2017ipp,Cheng:2017rdo} for tree, 2-dimensional convolutional neural networks (CNN)~\cite{deOliveira:2015xxd,Komiske:2016rsd,Kasieczka:2017nvn,Macaluso:2018tck} for jet image. 
Those delicate deep-learning approaches can better leverage the fine resolution of detectors and automatically figure out the complex pattern of a jet from the low-level inputs. 
However, those methods rely on traditional jet clustering algorithms to reconstruct the jet at the first stage. 
As a result, a predefined cone size parameter is required. And the jet representation could suffer from distortion due to inappropriate cone-size parameter or contaminations from pileup events. 
{The ParticleNet~\cite{Qu:2019gqs}, Particle Transformer~\cite{Qu:2022mxj}, LorentzNet~\cite{Gong:2022lye}, and PELICAN~\cite{Bogatskiy:2022czk} are among the state-of-the-art methods for Higgs and top tagging in this field. They achieve typical Area Under Curve (AUC) values of over 0.98 for top tagging, without considering the pileup effects. In addition, the momentum reconstruction component of PELICAN network can predict the $p_T$ and mass of $W$ boson with standard deviations of a few percent.}

Several studies~\cite{Grigoriev:2003tn,Mackey:2015hwa,Cerro:2021abp,Mukhopadhyaya:2023rsb} attempt to propose jet definitions alternative to the clustering methods, so that the presumed cone size parameter is no longer mandatory. 
Meanwhile, the techniques of object detection and semantic segmentation in computer vision provide new ways to label the jet constituents. 
In the Monte-Carlo simulation of collider events, the final state hadrons can be attributed to their ancestor parton without ambiguity. 
So it will be possible to build a neural network to label the jet constituents among final state particles based on supervised learning. 
Ref.~\cite{Ju:2020tbo} studies the construction of a $W$ boson jet from final state particles with the supervised GNN. 
In Ref.~\cite{Guo:2020vvt}, we improve the GNN with focal loss function, such that the method can remain efficient when heavy pileup contaminations are taken into account. 
Moreover, we demonstrate that the GNN, which is trained on events of the $H$+jets process, is capable of detecting a Higgs jet in events of several different processes. 
The image segmentation with the convolutional network also works well in detecting the Higgs jets in event images. 
In Ref.~\cite{Li:2020grn}, we take the event information as a digital image and adopt the Mask R-CNN framework~\cite{2017arXiv170306870H} to reconstruct the Higgs jet in the event image. 
Those deep learning methods reach higher efficiency of Higgs jet detection and higher accuracy of Higgs momentum reconstruction than the traditional jet clustering and substructure tagging methods. 

However, those methods have not been tested for detecting multiple jets of different kinds in an event. 
In this work, based on the event image representation, we adopt a modified version of Mask R-CNN to detect/reconstruct all Higgs and top quark jets in an event. 
Since the top quark is carrying a color charge, its energy flow is interconnected with other colored particles in the production process and with the beam remnants. 
There is no unique way to associate the hadronic final states with it.
Based on the rule that the vector sum of top jet constituents can reproduce the top quark momentum well, we propose a pattern of attribution for the top quark final states in the training sample. 
As a result, the Mask R-CNN can be trained to detect the top quark jet in a supervised way, similar to the Higgs jet. 
Moreover, the Mask R-CNN can only predict regions with masks which will also include a large number of particles from pileup events. The Higgs and top quark momenta can not be simply obtained by the vector sum of momenta of masked pixels (calorimeter cells) in the jet image. 
We add a new fully connected network component to the Mask R-CNN, which takes the input of bounding box information to predict the Higgs and top quark momenta. 
Given the Higgs and top quark momenta at the ground truth level, this network component is capable of pileup mitigation in an automatic way after supervised learning. 
It turns out that the modified Mask R-CNN can not only provide the jet regions (masks) of Higgs and top jets in an event but also predict the four-momenta of the Higgs and top parton precisely. 
{We compare the Higgs and top jet tagging performance with LorentzNet and the momentum regression performance with PELICAN.}

This paper is organized as follows.
In Sec.~\ref{events}, we describe the event generation and the event preprocessing. 
In Sec.~\ref{maskrcnn}, we briefly introduce the Mask R-CNN framework and illustrate how events proceed. 
The changes to the Mask R-CNN are discussed in detail.  
In Sec.~\ref{perform}, the performances of the network being applied to the $H t \bar{t}$ process as well as other processes that have not been used for training are presented. 
We summarise our work and conclude in Sec.~\ref{conclude}.

\section{Event preparation and preprocessing} \label{events}
The proton-proton collision events are simulated by the \texttt{MG5\_aMC@NLO} framework~\cite{madgraph} with center-of-mass-energy $\sqrt{s}=13$ TeV. The \texttt{Pythia8}\cite{pythia} is used for the quark parton showering, hadronization, and hadron decay. 
The detector effects are not considered except for the angular granularity of calorimeters~\footnote{The effects of energy smearing will be discussed separately later.}. The angular size of the calorimeter cell is assumed to be $0.02\times 0.02$ on the $\eta \times \phi$ plane. 
{This is an idealized setup since the hadron calorimeter at the LHC usually has $\eta/\phi$ resolution larger than $\sim 0.15$. Although the precision of momentum reconstruction is limited by the cell size, we find that the network performance, {\it i.e.} the Higgs/top detection efficiency, is barely changing with the cell size.}
The event image is built by presenting each calorimeter cell on the $\eta \times \phi$ plane as a pixel of an image, and the transverse momentum of the cell as the intensity (or grayscale color) of that pixel. 

The network is trained on one million events of the $Ht\bar{t}$ process, where both the Higgs and top quark jets are marked. 
{It should be noted that training our network on other processes is certainly possible and the performance of the network should be similar.}
To show the generality of the network, the performances on various test samples are studied, including $Ht\bar{t}$, $t\bar{t}t\bar{t}$, $HHt\bar{t}$ production in the SM model as well as the neutralino pair production and top squark pair production with subsequent decay $\tilde{\chi}^{0}_{2} \to H \tilde{\chi}_1^0$ and $\tilde{t}\rightarrow t\tilde{\chi}_1^0$ in the supersymmetric (SUSY) model. 
The transverse momenta of Higgs and top quark in the SM processes are required to be greater than 200 GeV and 300 GeV, respectively. As for the SUSY case, we set the masses $m_{\tilde{\chi}^{0}_{2}}=450$ GeV, $m_{\tilde{t}}=650$ GeV and $m_{\tilde{\chi}^{0}_{1}}=100$ GeV. 
We do not specify the decay modes of the Higgs and top quark in the training sample. However, we find the network exhibit better performance on events with hadronically decaying Higgs ($H \to b \bar{b}$) and top quark ($t \to b W, W \to q q$). Therefore the Higgs and top quark in the test sample are forced to decay through those modes. 

Moreover, there are multiple proton-proton collisions (referred to as pileup) in each bunch crossing at the LHC. Those collisions are dominated by nondiffractive events with small transverse momentum transfer. 
Simulation of the pileup events requires perturbative parton shower, Lund-string hadronization, multiple parton interaction and colour reconnection, which are usually described by phenomenological models. 
The parameters in the models are not unique and need to be inferred from experimental data. 
The set of appropriately chosen parameters is dubbed Pythia tunes~\cite{ATLAS:2012uec}. 
We adopt the A3 tune of \texttt{Pythia8} with phenomenological parameters provided in Refs.~\cite{Skands:2014pea,ATLAS:2016puo} to simulate pileup events.
The number of pileup events per bunch crossing at the LHC follows the Poisson distribution with an average value around  $\langle \mu \rangle$ = 35 at the LHC run-II and $\langle \mu \rangle$ = 200 at High-Luminosity LHC. 
We took the average number of pileup events of $\langle \mu \rangle$ = 50 in our simulation. {A detailed study on the effects of different pileup levels will be given later. }
Finally, we note that particles flying into the same calorimeter cell can only be identified with the summation of their momenta. So, the pileup will increase the momenta of the target jet constituents, and the pileup mitigation procedure is essential to obtain a precise parton momentum.

\subsection{Attributions of final state particles} 
In the training sample, the constituents of the Higgs jets and top/anti-top quark jets need to be assigned beforehand. However, due to color confinement, some of the top quark final states could have multiple ancestors other than the top quark according to the Monte-Carlo simulation. In constructing the top jet, we hope to only include the constituents whose momenta are mostly inherited from the top quark. 

\begin{figure}[thb]
\includegraphics[width=0.48\textwidth]{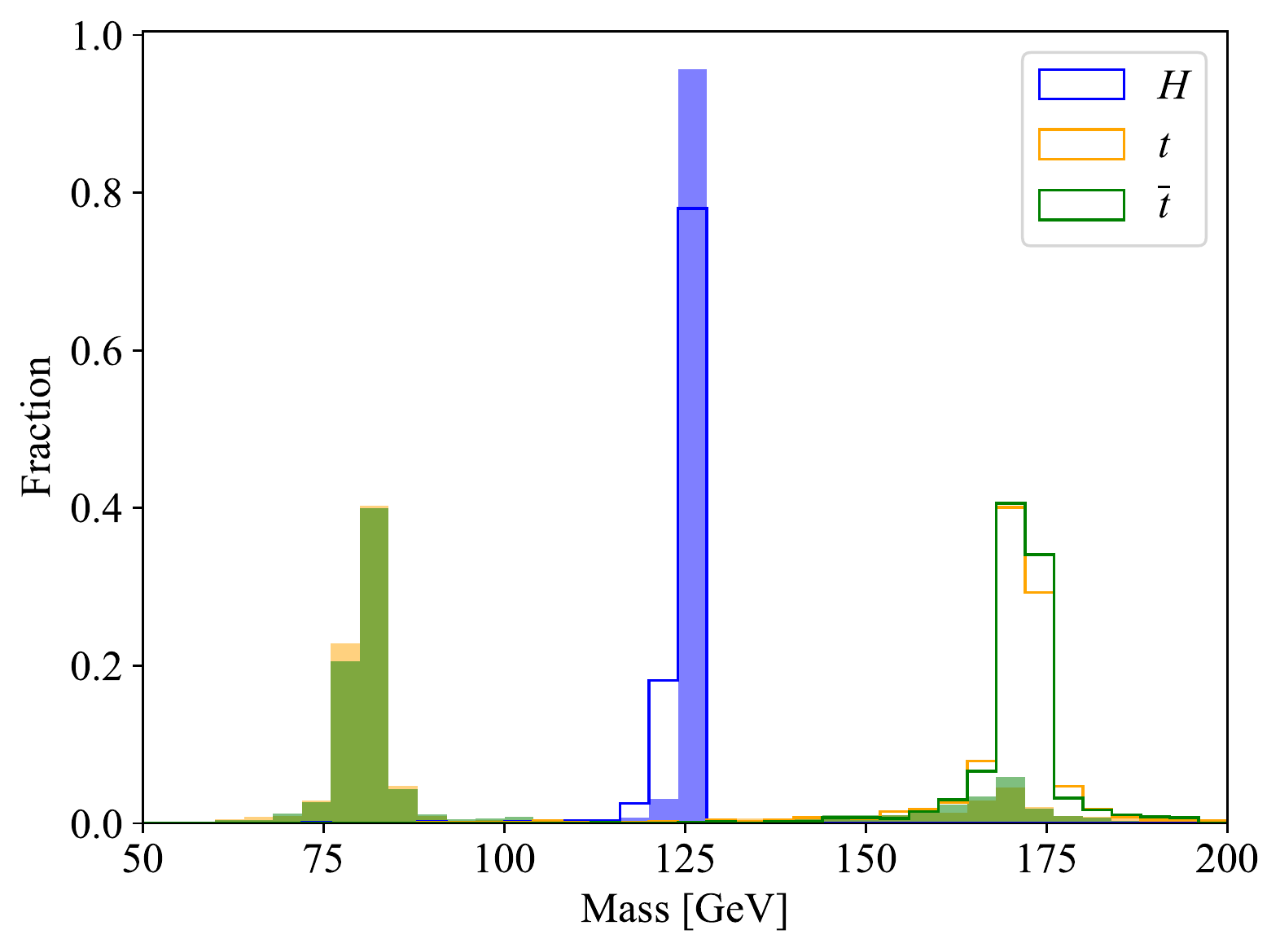}
\includegraphics[width=0.48\textwidth]{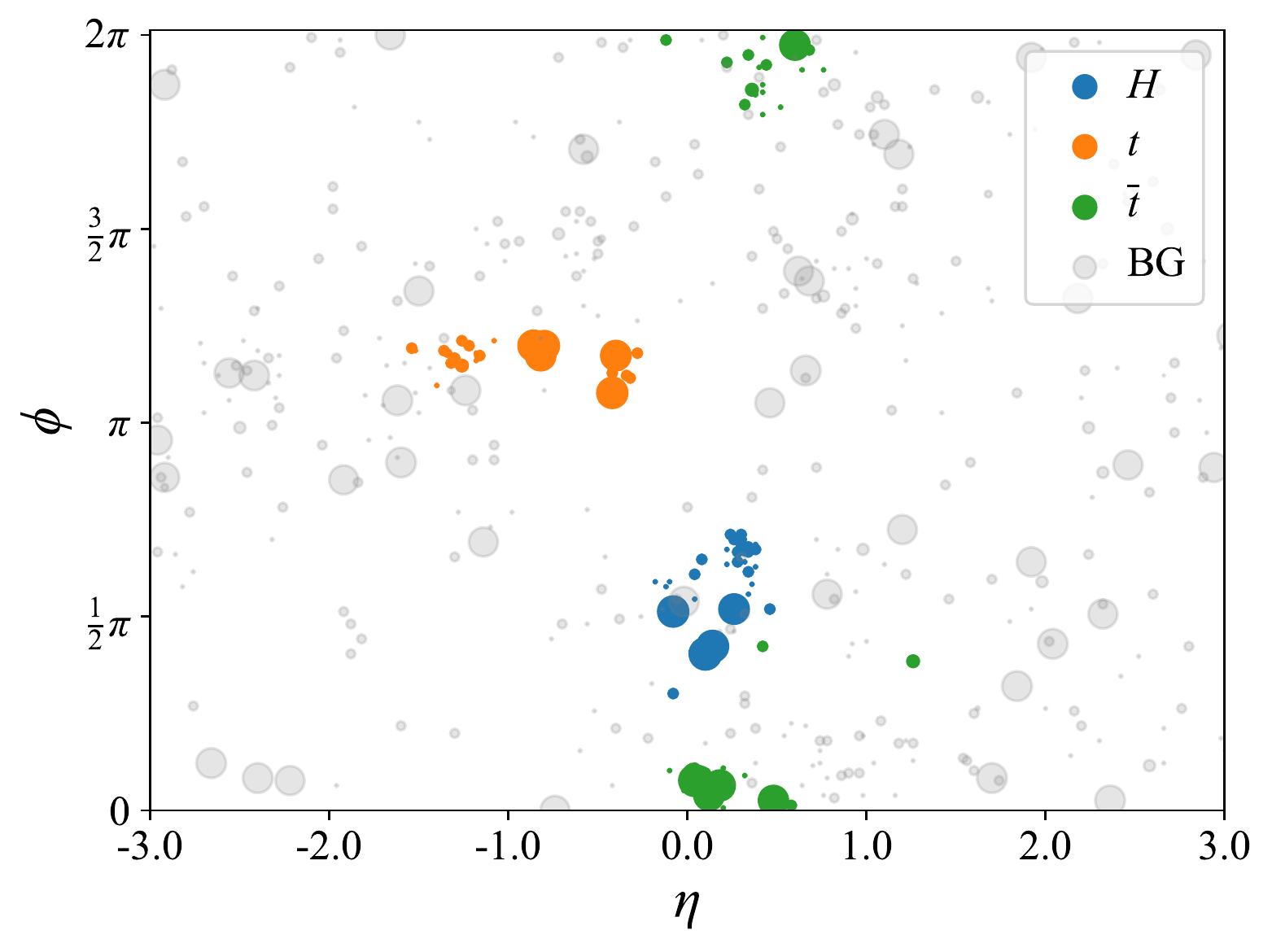}
\caption{\label{fig:bfa}
Left: distributions of the invariant masses for Higgs jet, top quark jet, and anti-top quark jet in events of $Ht\bar{t}$ production at the LHC. The shaded histograms and the solid lines correspond to the distributions before and after including the selected hadrons in the MA category. Right: the event image after applying the assignment, where the size of each dot indicates the energy of each final state hadron. 
}
\end{figure}

The final states of an $Ht\bar{t}$ event fall into four categories. Those who only have a unique ancestor should be assigned to $H$, $t$, and $\bar{t}$ categories without ambiguity. 
The rest of the final states have multiple ancestors (dubbed as MA category) and should be assigned to top/anti-top with some criteria. 
Hadrons in the MA category (one of the ancestors is the top quark or anti-top quark) are ranked according to their angular distances ($\Delta R = \sqrt{\Delta \phi^{2} + \Delta y^{2}}$) to the top/anti-top quark, where $\phi$ is azimuth angle, and $y$ is rapidity. 
They are assigned to the top/anti-top categories in order until the reconstructed top/anti-top jet invariant mass exceeds 1.05$m_{t}$ with $m_{t}$ being the top quark invariant mass (which could be off-shell) in the event~\footnote{Although this criterion can not guarantee the total match between the momenta of the top quark and the top jet, this method already help to capture most of the high energy constituents. Modifying the criterion will change the predicted mask (or jet shape), but will not have significant effects on the predicted momenta.}. 
In the left panel of Figure~\ref{fig:bfa}, we show the distributions of the invariant masses for the Higgs jet, top quark jet, and anti-top quark jet before and after the assignment. 
In terms of the invariant mass, we could observe that the selection of hadrons in the MA category is necessary, and our assignment criterion is appropriate. In the right panel of the same figure, we illustrate the event image on the pseudorapidity ($\eta$) versus the azimuth angle ($\phi$) plane after applying the assignment. Both the top and anti-top jets have focused shapes since their constituents are selected according to the angular separation. 

\subsection{Data preparation for the network}

The images fed to the network represent the transverse momentum 
deposited in the $\eta \times \phi$ plane.
Since the pixel values of typical images are integers ranging from 0 to 255 in each channel,
one may consider mapping from $p_T$ to the integer values in RGB.
However, we instead use $p_T$ itself
as we find no disadvantages in the network performance.
When the range of $\phi$ in the images is from 0 to 2$\pi$,
the constituents of a jet with $\phi$ near the boundary ($\phi = 0$)
may locate in two regions far apart from each other,
which are susceptible to being considered as two different objects.
In order to incorporate the periodicity in the network,
one may introduce a periodic padding in convolutional layers,
where an input image is padded in the periodic manner so that convolution kernels 
can read the values on the opposite side of the boundary.
Another way to handle the periodicity is
to have the images with an augmented range, for instance, $\phi \in [0, 3\pi]$,
such that a continuous picture of a jet appears on the image at least once.
In our brief implementation of the two schemes,
we observed better performance from the augmented images.
Therefore, we use the augmentation scheme in this paper.
Considering together network requirements for image dimensions,
the $\phi$ range is chosen to be from 0 to $2.85 \pi$.
The spatial size of input images is then $448 \times 448$ pixels
corresponding to $\eta \times \phi$ plane across $[-4.48, 4.48] \times [0, 2.85 \pi]$
where the resolution is given as $\Delta \eta \times \Delta \phi =  0.02 \times 0.02$.
Having three copies along the channel,
the dimension of the input images is $448 \times 448 \times 3$.
{Having three copies is not just redundant due to the network requirement, but it has a non-trivial effect on the network. It implies that the kernel at the first convolutional stage should also have three channels, tripling the number of learnable parameters.}

\begin{figure}[htb]
\centering
\includegraphics[width=1.0\textwidth]{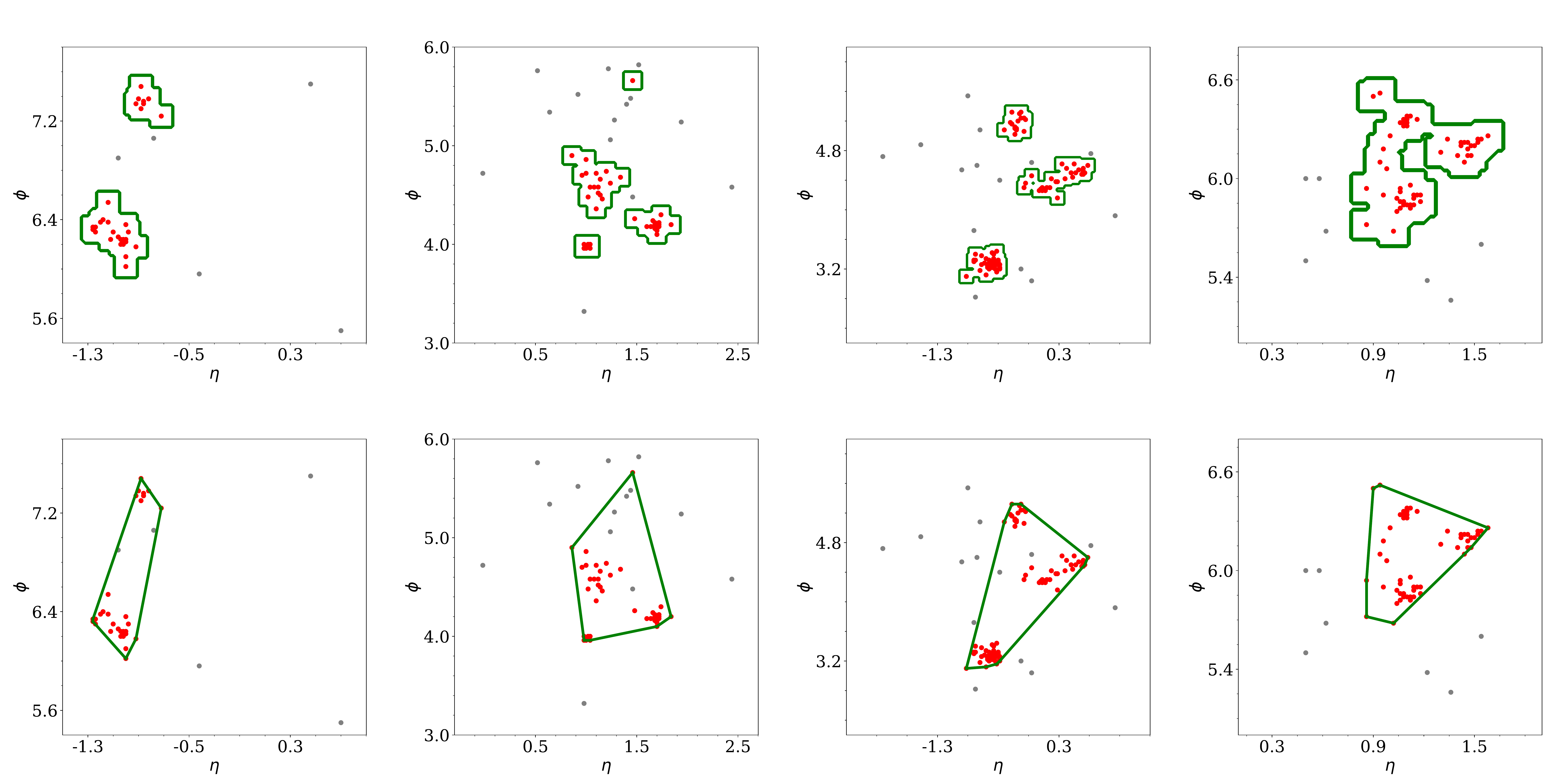}
\caption{\label{fig:mask_ex}
Top:  enlargement areas. Bottom: convex hull areas.
Red dots are the constituents constructing the jet areas represented by the green lines.
Grey dots are the ones ruled out by the neighbourhood condition.
One may see that, without the condition, 
how drastically few grey dots could change jet areas.
Nevertheless, we spare a constituent with high $p_T$ 
($>$ 5 GeV)
as shown in the second example (the red dot on the top).
}
\end{figure}

Given an input image, 
the original Mask R-CNN has three outputs for each candidate object,
a class label, a bounding box, and a mask.
Therefore, one needs to provide
the ground truth of them to train the network,
but there is an issue to address here.
The mask is a binary image with the same spatial size as the input image,
in which object pixels have the value 1
and background pixels have 0,
and the smallest rectangle enclosing all the object pixels is the bounding box.
In the jet detection task,
it is reasonable to define
the pixels where individual jet constituents are located
as the object pixels.
Note that, however,
Mask R-CNN will crop candidate regions in the mask
and resize them into a fixed size 
to compare to corresponding outputs of the network.
The sparsity of the constituent pixels is not robust to resizing, 
in particular, when scattered in a broad region.
Instead, we define a jet area that gives a mask of connected pixels.
First, we preselect constituents which will compose the jet area.
The preselection process is as follows.
Boost along the beam direction to the frame
where $p_z$ of the parton is zero,
and discard the constituents with energy lower than 0.1 GeV
or with angular separation to the parton greater than $\pi / 2$.
Then, among the remaining constituents,
select those having at least 3 others
in $20 \times 20$ pixels neighbourhood
or with $p_T$ greater than 5 GeV.
The neighbourhood condition is imposed
since we want to construct jet areas
that do not change drastically
depending on a couple of constituents
unless $p_T$ is significantly large.
Otherwise, the bounding box predictions
will fluctuate widely according to whether or not
the network precisely detects a few constituents far from some clusters.
The size of neighbourhood and the number of neighbours
are empirically determined after monitoring
the network performance of several cases.
With the selected constituents,
we define two types of jet area, the convex hull and the enlargement.
The first one is the area 
bounded by the convex hull covering all the selected ones,
and the second one is obtained
by expanding each selected pixel into the area of $9 \times 9$ pixels
with the selected one at the center.
Now we define each pixel in the jet area as the object pixel.
The convex hull mask is a simply connected region,
as is usual for object masks in natural images,
whereas the enlargement mask may consist of several regions
useful to identify clusters of constituents more precisely (Figure \ref{fig:mask_ex}).

In natural images, 
the bounding box
is an intuitive and also robust concept since
objects have their boundary,
and the deviation of the bounding box
depending on specific choices of boundary pixels is not large.
On the contrary,
there is no such thing as a boundary for jets,
and instead we introduced the jet area
in order to alleviate the issues of the sparse object pixels
and the large fluctuation of the bounding box.
Nevertheless,
the size and shape of 
the bounding box will rely on our preselection rules.
It may be helpful to adjust the bounding box 
by involving a ground truth value 
irrelevant to our definition of jet areas.
Therefore, we employ the ground truth coordinates of partons,
i.e., the rapidity $y$ and azimuth $\phi$,
to be the center of the bounding box.
The bounding box is now the smallest rectangle enclosing the jet area
while its center is fixed at the ground truth coordinates of the parton
\footnote{
Since the approximation $y \simeq \eta$ is valid for ultrarelativistic particles,
the coordinates of a parton on the $\eta \times \phi$ plane
should take its $y$ and $\phi$.}.
We expect that the adjustment tells the network consistent information
about the keypoint of a jet around which its features should be extracted and gathered
through convolutions
regardless of our choice of preselection rules.
Indeed, we find a significant improvement in the network performance
compared to using the default bounding box.

\section{Mask R-CNN and its modifications} \label{maskrcnn}

Mask R-CNN \cite{8237584} is a state-of-the-art framework
for object detection and instance segmentation.
It was progressively developed 
from region-based convolutional neural networks (R-CNNs) 
that first take candidate regions
from separate region proposal methods,
and then use a convolutional network to extract features
for classifications and bounding box regressions.
The original R-CNN \cite{6909475} was computationally expensive
as it performs a CNN for each region proposal.
By extracting a feature map from the entire input image
to share across region proposals,
SPPnet \cite{10.1007/978-3-319-10578-9_23} greatly reduced the computational cost,
and Fast R-CNN \cite{7410526} streamlined a multi-stage pipeline of the predecessors
as the classifier and box regressor are jointly trained with the feature extraction network.
While the previous models take the region proposals
from separate region proposal methods,
Faster R-CNN \cite{NIPS2015_14bfa6bb} has brought them into one unified network
by introducing region proposal networks (RPNs)
that can share the convolutional feature map,
showing remarkable gains in speed and accuracy.
Finally, Mask R-CNN extends Faster R-CNN
by adding a mask branch for instance segmentation
in parallel with the classifier and box regressor.

\begin{figure}[tpb]
\centering
\includegraphics[width=1.0\textwidth]{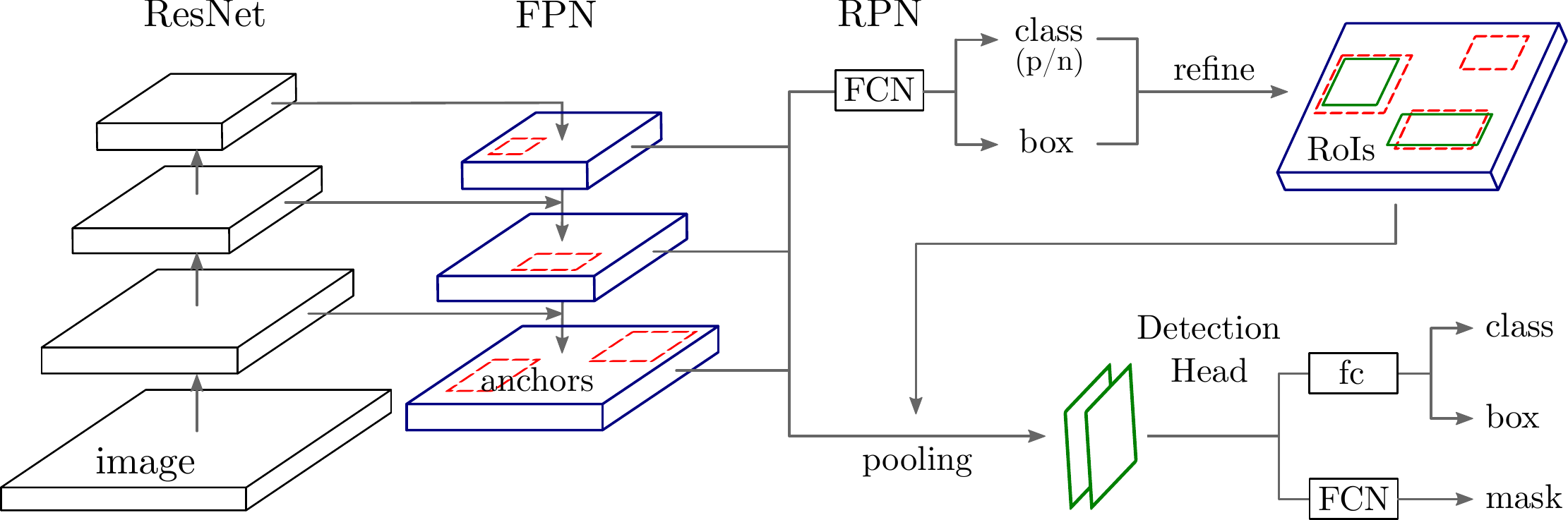}
\caption{\label{fig:MRCNN}
Schematic diagram of Mask R-CNN, adapted from \cite{Li:2020grn}.
The backbone shows only three convolutional stages of ResNet
instead of five for simplicity.
FCN denotes a fully convolutional network, and
fc represents fully connected layers.
}
\end{figure}

Let us briefly review the structure of Mask R-CNN (Figure \ref{fig:MRCNN}).
It can be mainly divided into three modules,
a backbone architecture for feature extraction,
an RPN for region proposal generation,
and a detection head for classification, box regression, mask segmentation.
The RPN and detection head share the backbone such that
the features are used for both regional proposals and detections.
Although the backbone can be any convolutional architectures,
Feature Pyramid Network (FPN) \cite{8099589}
is commonly used to take advantage of multi-scale features
in addition to the base architecture such as residual neural networks (ResNets) \cite{he2016residual}.
The ResNet extracts features from an input image
by successively scaling down its spatial size through convolutions,
which in turn produces a pyramid of outputs at several scales.
The FPN then uses a top-down pathway
with lateral connections to the outputs at each scale
in order to build a feature pyramid.
The feature map at each level of the pyramid is fed into the RPN
to propose candidate bounding boxes, 
referred to as regions of interest (RoIs).
An anchor is a reference box whose center is at a pixel of the feature pyramid.
By default every pixel has three anchors with aspect ratio of 1:1, 1:2, and 2:1,
where the anchor scale is different according to the level of the feature pyramid.
Through a small convolutional network,
the RPN outputs an objectness score and box regression for each anchor.
It is trained so that an anchor has a high score
if its Intersection over Union (IoU) with a ground truth box greater than a threshold,
and the box regression refines
the size and location of positive anchors
to fit their ground truth boxes better.
The refined positive boxes are the RoIs
that are passed on to RoI Pooling.
While the RoIs have variable sizes due to the refinement in the RPN,
the detection head requires a fixed-size input.
Therefore, we pool RoI features of a fixed size
by cropping the RoIs from the feature pyramid
and resizing them using bilinear interpolation
\footnote{
We use the crop and resize operation following the source code \cite{matterport_maskrcnn_2017} for simplicity.
The original paper \cite{8237584} proposed a more elaborate method, called RoIAlign,
to reduce misalignment between the RoIs and the extracted features.}.
The RoI features are then fed to the detection head that has two branches.
The classification branch consists of two fully-connected layers 
followed by classification and box regression outputs.
The mask branch is a small fully convolutional network
to predict binary masks.

\begin{figure}[h]
\centering
\includegraphics[width=0.6\textwidth]{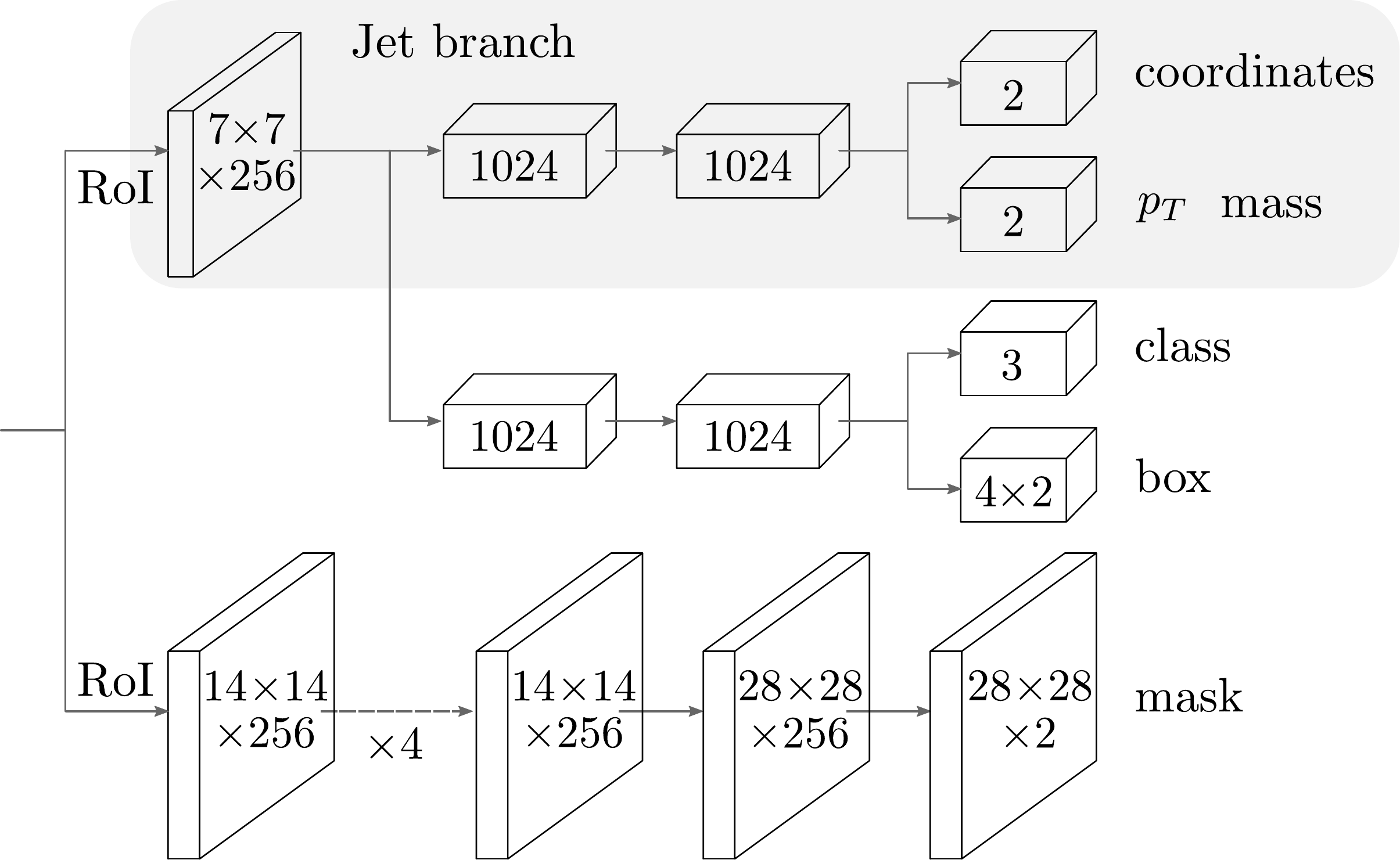}
\caption{\label{fig:head}
Detection head together with the jet branch.
Numbers inside boxes denote dimensions.
`$\times 4$' represents four consecutive convolutions.
The figure of classification and mask branches are adapted from \cite{8237584}.
}
\end{figure}

In the jet reconstruction task,
the original Mask R-CNN can be used to tag jets 
and predict the jet areas.
On the other hand,
the jet areas, especially the convex hull areas,
are susceptible to the pileup contamination
since they include background particles as well.
Furthermore, the preselection process to define jet areas
may exclude some constituents that significantly contribute
to the four-momentum of their parton.
Therefore, 
in order to accurately obtain four-momenta of partons 
from jet area predictions,
one needs separate methods 
carrying out the pileup mitigation
as well as compensating for the preselection.
On the contrary, instead of using separate methods,
we bring a pileup mitigation and compensation network into Mask R-CNN
to facilitate end-to-end jet reconstructions.
In other words, we extend Mask R-CNN
by adding an additional branch for predicting the mass, $p_T$, and coordinates of partons~\footnote{In practice, the vector sum of the constituents momenta is used as ground truth instead of the momentum of the original parton in training our network, because vector sum is more directly related to the masked constituents. However, in the training sample, the momentum difference between the vector sum and the parton is less than 2\% for  Higgs and 5\% for top, for more than 90\% events.},
which we call jet branch.
The jet branch has the same architecture as the classification branch,
i.e., two fully-connected layers, and also shares the RoI inputs
with the classification branch (Figure \ref{fig:head}).
It is worth noting that 
we predict two boxes and two masks for each RoI
(denoted by `$\times 2$' in the figure),
although it is usual to predict one box and mask per class
so that the default number of predictions for each RoI is $3$
corresponding to the three classes (background/Higgs/top).
If objects in each class have their typical shape or ratio,
the class-specific prediction may be more effective.
However, the Higgs jets are indistinguishable from top jets
by their bounding box or mask.
Therefore, we predict for two classes (background/jet)
considering Higgs and top as one class.
On the other hand, 
the jet branch outputs a single four-momentum
(coordinates, $p_T$, and mass) for each RoI regardless of class.
This reflects the fact that it is always possible to calculate
the four-momentum even for a background region in a consistent way.
Furthermore, this approach also can tell the network that
the mass of an RoI is essential to determine its class.
Consequently,
the mass and class prediction tasks will be closely intertwined
and enhance each other.

In our implementation of Mask R-CNN,
which is based on the open-source code in \cite{matterport_maskrcnn_2017}, 
we also employed Composite Backbone Network (CBNet) \cite{Liang_2022} as our backbone
and Cascade R-CNN \cite{8578742} as an extension of the detection head,
so as to further improve accuracy.
We present a brief introduction to the two architectures
while referring readers to the original papers \cite{Liang_2022, 8578742} for details.
The CBNet groups multiple identical backbones 
by connecting them in parallel.
We use CB-ResNet50 which consists of
two ResNet50s, an assisting one and a lead one,
connected in such a way that
the features of higher-level stages in the assisting backbone
flow to the lower-level stages in the lead backbone.
Therefore, the lead backbone can integrate the high-level features
into its low-level convolutional stages
for more effective feature extractions.
Cascade R-CNN extends the detection head
in order to have more accurate bounding box predictions.
The detector requires an IoU threshold
to decide whether an RoI is positive or negative,
and the commonly used threshold value is 0.5,
which will be robust to poor proposals
but also can be loose,  leading to noisy box predictions.
To address the problem,
Cascade R-CNN  adds detectors
and constructs a sequence of detectors
with increasing thresholds at each stage.
We use three detectors
where the first two have only the classification branch
to refine RoIs while the last detector has all three branches.

\section{Network performance} \label{perform}

\begin{figure}
\centering
\includegraphics[width=0.9\textwidth]{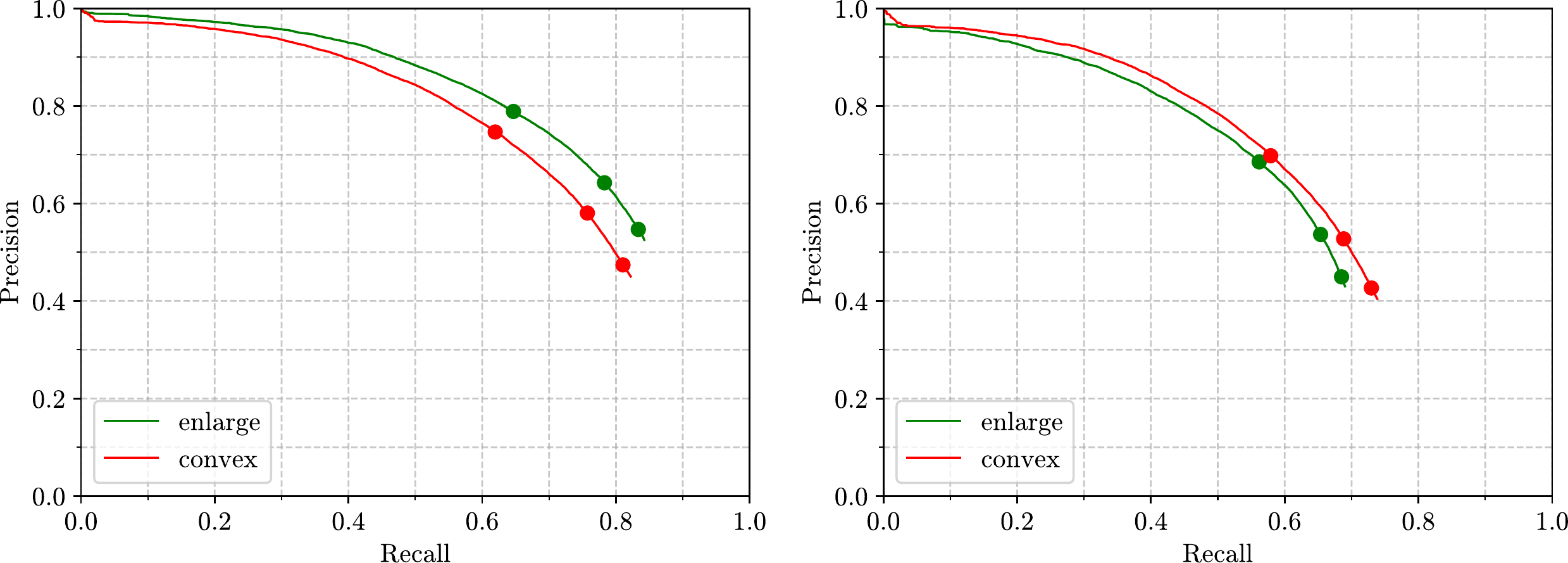}
\caption{\label{fig:PRcurve}
Precision-Recall curve at IoU=0.5 
for the bounding box (left) and mask (right).
Three dots on each curve indicate the points
where the score threshold is 0.9, 0.7, and 0.5 from left to right.}
\end{figure}  

\begin{figure}[htb]
\centering
\includegraphics[width=1.0\textwidth]{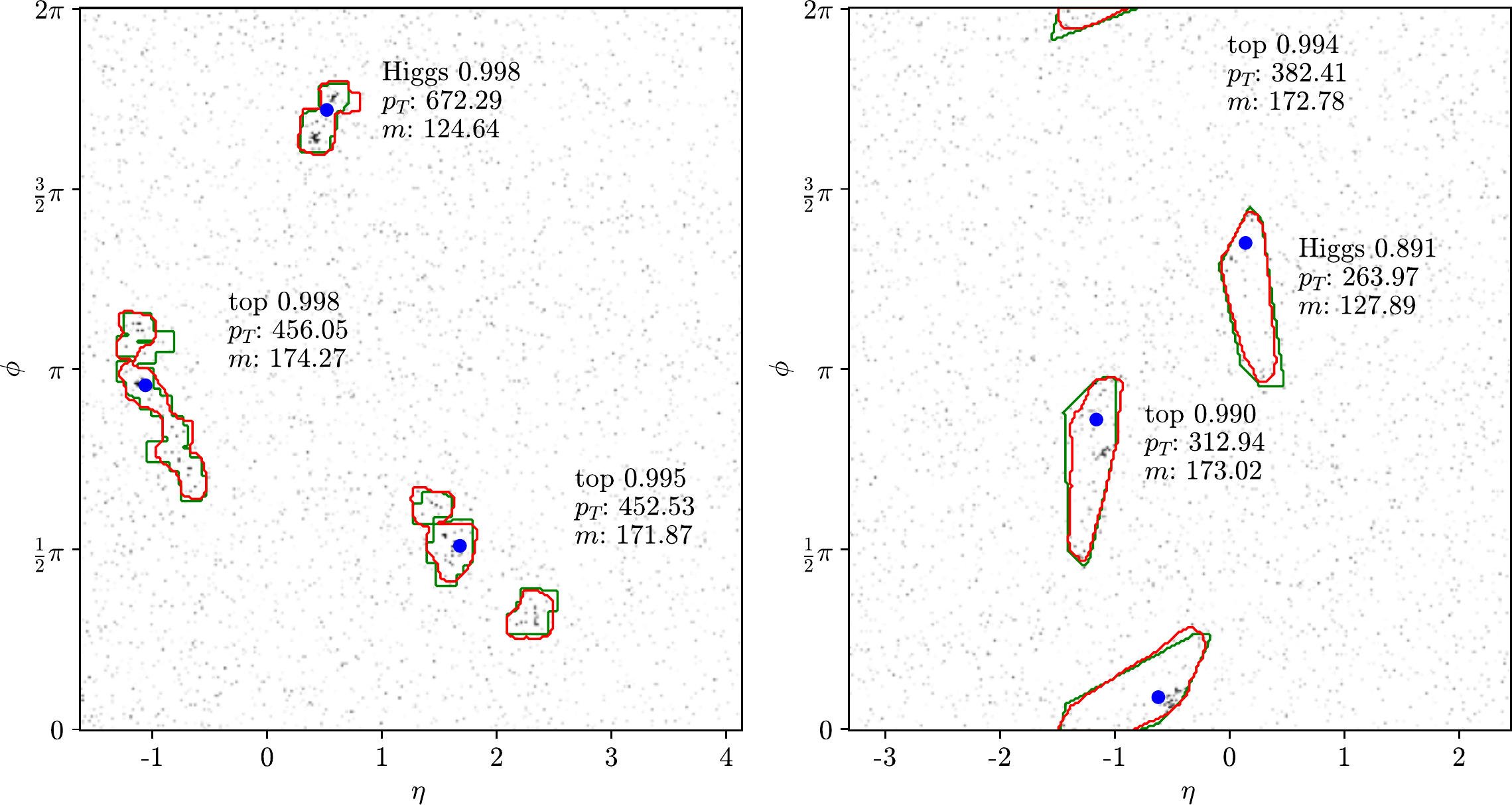}
\caption{\label{fig:enlarge_convex_result}
Prediction examples.
Left: enlargement mask.
Right: convex hull mask.
Green and red lines represent ground truth and predicted masks, respectively.
Labels include score, $p_T$, and mass predictions.
Blue dots denote predicted coordinates.}
\end{figure}

In object detection, it is usual to evaluate the network performance
using  Average Precision  (AP).
Given an IoU threshold,
one can obtain the precision-recall curve by varying the score threshold
as shown in the Figure \ref{fig:PRcurve},
and AP is the area under the curve.
In jet detection, however,
the value itself is not directly comparable to that of other models
since AP also heavily depends on the mask scheme.
Nonetheless, it is a useful metric within one mask scheme,
and hence we report 
the mask AP and box AP in Table~\ref{tab:AP}
to show the performance in jet area detection
(see also Figure \ref{fig:enlarge_convex_result} for example).

\begin{table}[h]
\centering
\begin{tabular}{l | c c | c c}
		&  $\mathrm{ AP^{mask}}$ & $\mathrm{ AP^{mask}_{50}}$  
		& $\mathrm{ AP^{box}}$ &   $\mathrm{ AP^{box}_{50}}$  \\  \hline 
convex hull   & 24.7 &  60.7 & 37.1 & 68.9 \\ 
enlargement  & 16.0 &  56.6 & 43.0 & 73.5 \\ 
\end{tabular}
\caption{AP(\%) on $H t \bar{t}$ test set. 
$\mathrm{AP_{50}}$ denotes the AP at IoU=0.5, and
AP denotes the AP averaged across IoU from 0.5 to 0.95
with a step size of 0.05.}
\label{tab:AP}
\end{table}

On the other hand, 
the main goal of the jet reconstruction task is to calculate
the four-momentum of partons, which are independent of the mask scheme.
Therefore, we want to measure how accurately
our extended Mask R-CNN can predict the mass, $p_T$, and coordinates.
In experimental analyses, we have prior knowledge about the type and number of jets in selecting the signal events.
Let us assume we have three ground truth Higgs jets 
on an image for illustrative purposes.
First recall that, since we use the augmented images,
there may be two predictions for the same jet
duplicated along the augmented $\phi$ coordinate.
To sort this out, we post-process the output
by taking it back to the $\phi$ range of one period and eliminating
one with a lower score
if two predictions are of the same class and their mask overlap is large.
After the postprocessing,
we choose three predictions of Higgs with the highest scores
if there are more than three Higgs predictions.
Otherwise, we take all available Higgs predictions.
A prediction is a true positive if the distance between 
the predicted and ground truth coordinates 
is less than a threshold (we use 30 pixels $\simeq$ 0.6).
We measure the coordinates, mass, and $p_T$ differences
between the true positive prediction and its ground truth {(obtained by the vector sum of the constituents momenta)}.

\begin{figure}[htb]
\centering
\includegraphics[width=0.45\textwidth]{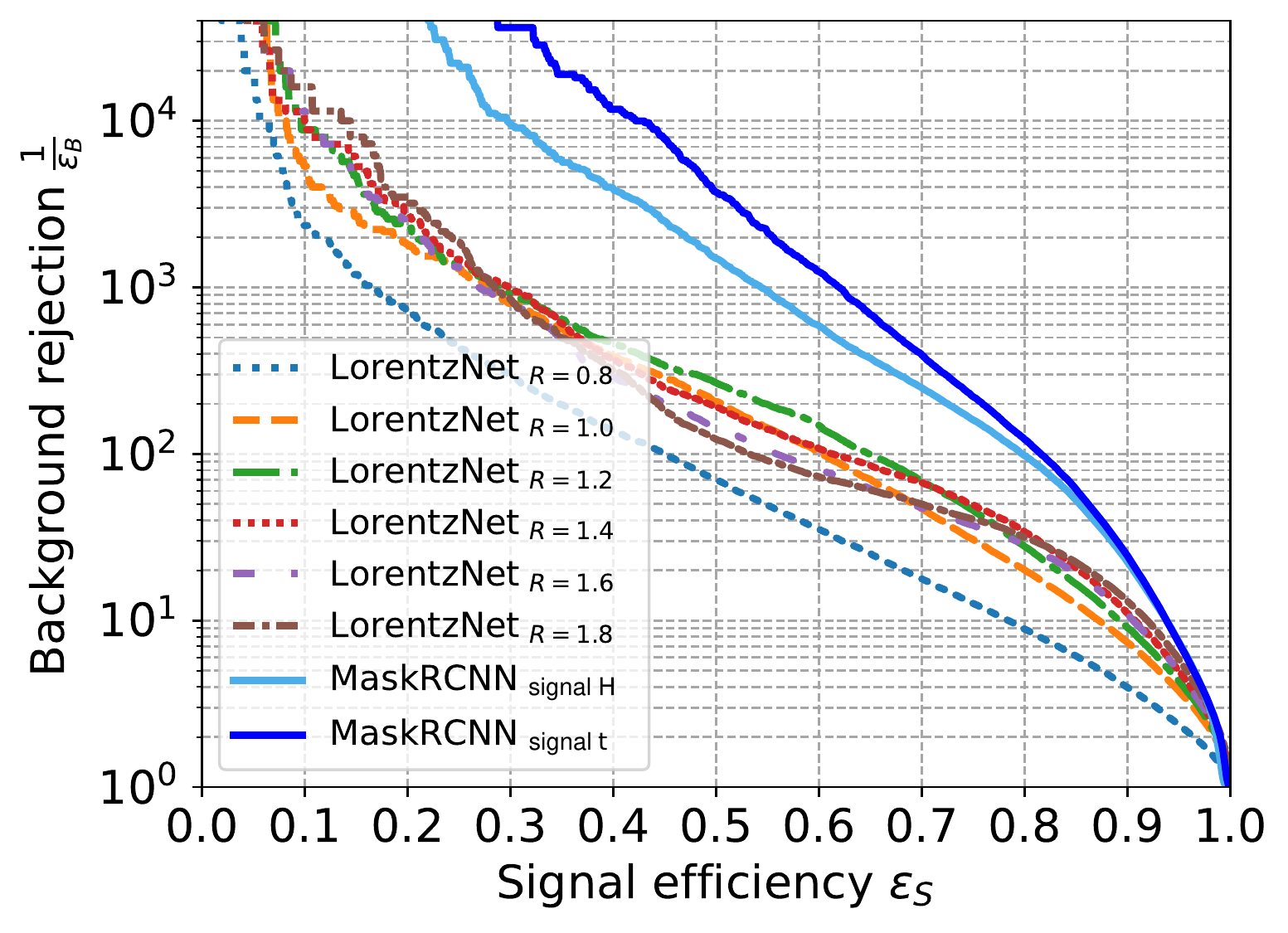}
\includegraphics[width=0.42\textwidth]{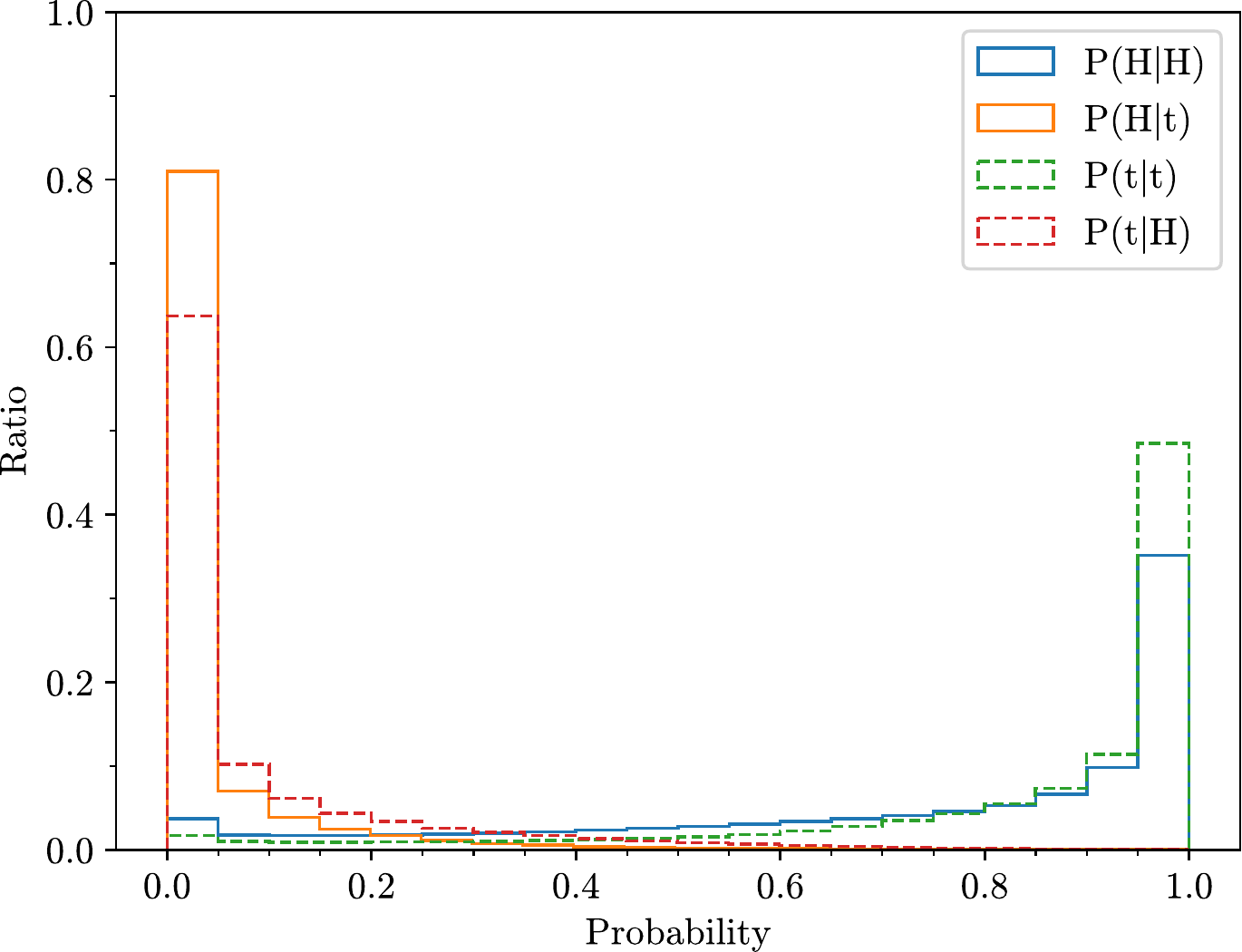}
\caption{\label{fig:complp} Left: ROC curves for the LorentzNet with different cone size parameters (top quark jet is taken as the signal and Higgs jet is the background) and for the Mask R-CNN (signal Higgs and background top in cyan color; signal top and background Higgs in blue color). Right: the distributions for the classification scores in Mask R-CNN. }
\end{figure}

{To gain an intuitive understanding of the network performance, we contrast our results with those from existing state-of-the-art object tagging architecture. 
We adopt LorentzNet for classification and PELICAN for momentum regression as an illustration. 
The validations for the application of these two networks are provided in Appendix~\ref{app:val}. 
These networks require a jet clustering algorithm to localize jets for a given event image, as they use a jet as the input. 
{We retrain these networks on Higgs and top quark jets from our 0.3 million $Ht\bar{t}$ event sample (with an average number of 50 pileup events superposed on each signal event), where the Higgs ($p_{T_H}>$ 200 GeV) and top quark ($p_{T_t}>$ 300 GeV) jets are reconstructed by the anti-$k_T$ algorithm with cone size parameter $R$ varying from 0.8 to 1.8 in steps of 0.2.
The detector effects have been ignored in this comparison study. 

It should be noted that the LorentzNet and PELICAN follow a fixed input format: a set of four momenta of the jet constituents. To simplify the training, we construct a training file for the reconstructed anti-$k_T$ jets mentioned above, mainly containing the following information:

\begin{itemize}
\item Nobj: The count of jet constituents.
\item Pmu: The four-momentum of jet constituents $(E,p_x,p_y,p_z)$, sorted in descending order of $p_T$. This part serves as the network input, with a shape of [N$\times$4], where N(=200) is the maximum number of jet constituents for a single input, and the insufficient parts are padded with zeros.
\item label: Indicates whether Pmu at a certain position is a constituent(1) or a padding value(0), with a shape of [1$\times$200].
\item truth\_Pmu: The four-momentum of the parton to which the jet belongs, used for momentum prediction.
\item is\_signal: The type of the jet determined by the distance from jet to parton. For example, if it is closest to the Higgs (i.e., when $\Delta R=\sqrt{\Delta\phi^2+\Delta y^2}$ is the smallest), it is classified as a Higgs jet(0), and top jet(1) is in a similar situation. This value is used for jet classification.
\item mass: The invariant mass of the parton to which the jet belongs.
\end{itemize}
The above file format refers to \url{https://zenodo.org/record/7126443}, and there are some other configurations mentioned in the same link, which are irrelevant to the training. For more details, please refer to the link provided above. The hyperparameters for both networks are set to the default values as shown in Refs~\cite{Gong:2022lye,Bogatskiy:2022czk}.}

The receiver operating characteristic (ROC) curves of the LorentzNet for top quark (signal) and Higgs (background) jet discrimination with different cone sizes are shown in the left panel of Figure~\ref{fig:complp}. 
It can be observed that the cone size parameter has non-negligible effects on the performance of LorentzNet. In particular, the performance degrades dramatically when the cone size becomes too small to fully capture the jet constituents. On the other hand, a larger cone size does not impair jet classification much, despite distorting the jet shape.
The performance of those classifiers 
is not directly comparable to that of Mask R-CNN.
However, we may consider only the classification part of Mask R-CNN
to obtain ROC curve.
In Mask R-CNN, RPN does a similar job to jet clustering algorithms.
The RPN in our hyperparameter setting proposes up to 500 RoIs for each image
\footnote{Overlaps between RoIs are allowed as long as IoU is less than a threshold (we use 0.7)
such that several RoIs indicate one jet in general.},
which will be passed to the detection head.
{To calculate the ROC, we first get all those RoIs of an event from RPN, and compute IoUs between RoIs and ground truth bounding box. The RoIs with IoU greater than 0.5 are selected and fed into the classifier branch. The RoI with the highest classification score is chosen for calculating the ROC curve of the Mask R-CNN. 
}
As our network predicts three classes,
we provide two ROC curves, one of which
considers Higgs jet as positive class and top quark jet as negative class,
while the other considers the opposite.
The ROC curves are presented in Figure~\ref{fig:complp}.
The Mask R-CNN achieves higher performance than the LorentzNet in general, due to its more accurate jet boundaries. 
{Those features can be quantified in terms of AUC values, as given in Table~\ref{tab:auclm}. }
In addition, the Mask R-CNN can tag top quark jet more accurately than Higgs jet in its trinary classification. 
The right panel of Figure~\ref{fig:complp} illustrates this fact with the classification scores distributions, where $P(H|H)$, $P(H|t)$, $P(t|t)$ and $P(t|H)$ are the probabilities of tagging true Higgs as Higgs, true top as Higgs, true top as top and true Higgs as top, respectively. 
}

\begin{table}[htb]
\centering
\begin{tabular}{| c | c | c | c | c |c | c | c | c | }
\hline
          & \multicolumn{6}{|c|}{LorentzNet } & \multicolumn{2}{|c|}{Mask R-CNN} \\ \cline{2-9}
	  & $R=0.8$ & $R=1.0$ & $R=1.2$ & $R=1.4$ & $R=1.6$ & $R=1.8$ & signal Higgs & signal top \\ \hline
AUC & 0.920 & 0.953 & 0.960 & 0.964 & 0.962 & 0.966 & 0.9723 & 0.9754 \\ \hline
\end{tabular}
\caption{AUC values for LorentzNet with different jet cone sizes and for the Mask R-CNN with either Higgs or top being the signal. \label{tab:auclm}}
\end{table}

\subsection{The reconstruction accuracy for the test sample}

\begin{figure}[htb]
\centering
\includegraphics[width=0.85\textwidth]{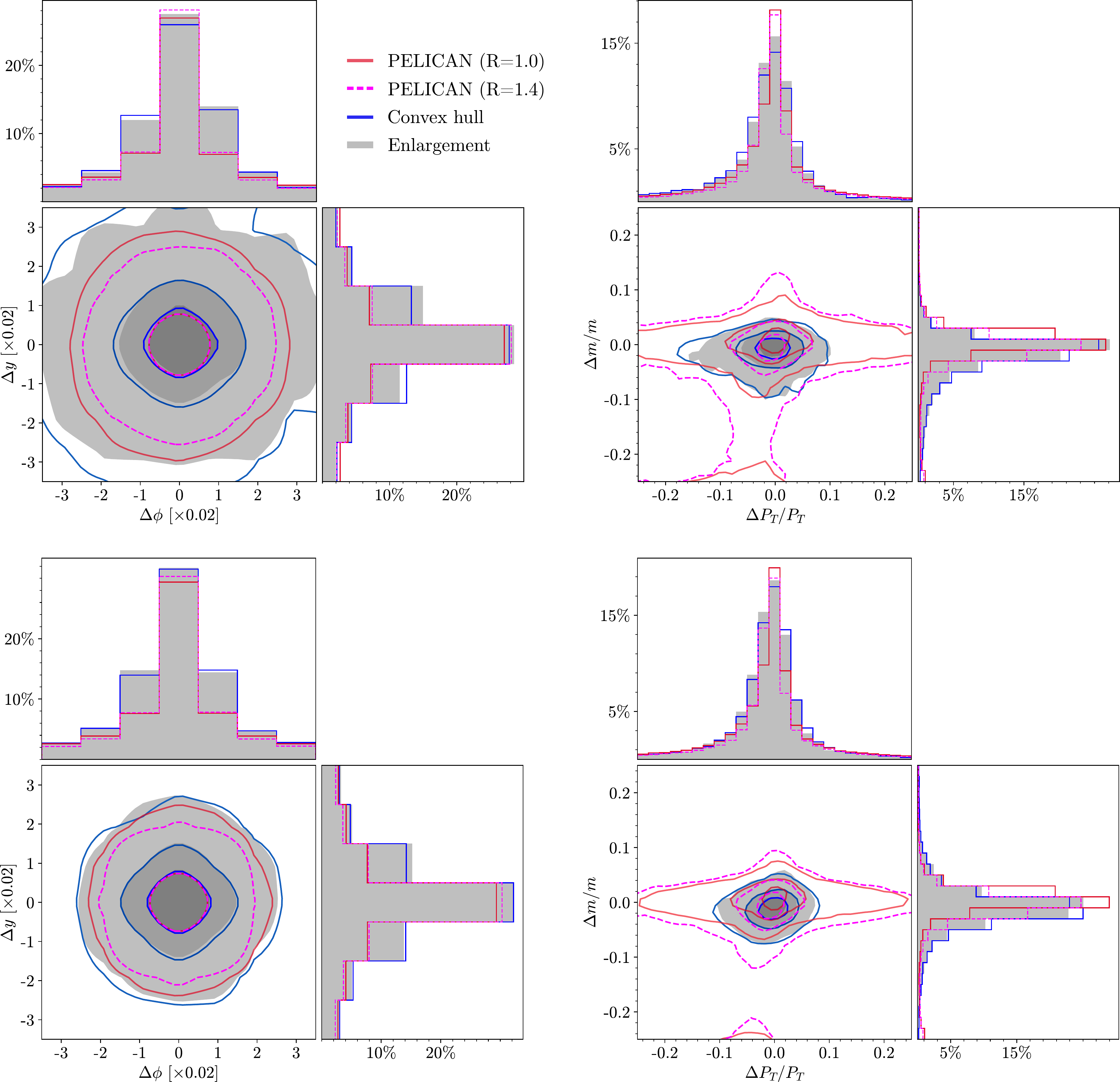}
\caption{\label{fig:httx_dist} The 2-dimensional event distributions of the $Ht \bar{t}$ event sample. Contours/shade from the inside out correspond to the event fraction of 20\%, 40\% and 60\%, respectively. 
Upper panels: reconstructed Higgs jets.
Lower panels: reconstructed top quark jets.}
\end{figure}  

The reconstruction accuracies of the Higgs and top four-momenta obtained from applying the PELICAN and the Mask R-CNN method to the test sample are illustrated in Figure~\ref{fig:httx_dist}, in terms of two-dimensional distributions on the $\Delta y \times \Delta \phi$ plane and $\frac{\Delta m}{m} \times \frac{\Delta p_{T}}{p_{T}}$ plane. Here $\Delta m, \Delta p_{T}, \Delta y$ and $\Delta \phi$ correspond to the differences of the invariant mass, transverse momentum, rapidity and azimuth angle between the reconstructed jet and the ground truth jet. 
{It should be noted that the four-momentum of the ground truth jet is obtained by the vector sum of its constituent momenta.}
The $m$ and $p_{T}$ in the denominator correspond to the values of the ground truth jet. 
The distributions are normalized such that the sum of all simulated events for each process is equal to one. 
In the figure, {the red and pink contours correspond to the distributions of the PELICAN method with cone size $R=1.0$ and $R=1.4$, respectively.}
The blue contours and grey shades correspond to the distributions of the Mask R-CNN methods with the convex hull mask and the enlargement mask, respectively. 
Different shades of gray regions and colored contours from inside out indicate 20\%, 40\% and 60\% of events, respectively. The closer they are to the center, the higher accuracy they stand for. 
It can be observed that the Mask R-CNN methods with different definitions of mask can achieve similar accuracies in both Higgs and top jet momenta reconstruction. 
For the test sample of the $H\bar{t}t$ process, about 60\% of Higgs jet can be reconstructed with $\Delta y \sim \Delta \phi \in [-0.06, 0.06]$, $\frac{\Delta m}{m}\in [-0.08,0.05]$ and $\frac{\Delta p_{T}}{p_{T}} \in [-0.14,0.1]$. And about 60\% of top jet can be reconstructed with $\Delta y \sim \Delta \phi \in [-0.05, 0.05]$, $\frac{\Delta m}{m}\in [-0.07,0.05]$ and $\frac{\Delta p_{T}}{p_{T}} \in [-0.11,0.08]$. 
{The Mask R-CNN method surpasses the PELICAN method in regressing both Higgs and top momenta for any choice of the cone size parameter, {when 60\% of most accurate events are considered}.  Note that the PELICAN efficiencies of successful Higgs and top tagging on $Ht \bar{t}$ events are around 78\% for $R=0.8$ and around 85\% for other cone sizes. } 

{To give a quantitative comparison between the performances of the Mask R-CNN and the PELICAN, we calculated the root mean squared distance (RMSD) using the following method. We note that using RMS in the usual way is not available since many of the predictions are false positives, which are outliers. 
From the prediction of two networks, we first sort the events by the distance of $(\Delta p_T/ p_T, \Delta m/ m)$ from the origin (0,0), {\it i.e.} sort the events according to the $p_T$ and $m$ prediction accuracies. 
Then, we calculate the RMSD for 10\%, 20\%, 30\%, 40\%, and 50\% of the most accurate events. 
The results of the test sample are given in Table~\ref{tab:rmsdhtt}. 
In the sense of RMSD, the fraction of events when Mask RCNN surpasses PELICAN is $35\%$ for Higgs momentum reconstruction, and is $40\%$ for top momentum reconstruction, respectively. 
In other words, those two methods have the same RMSD on Higgs (top) momentum reconstruction when the first 35\% (40\%) of events with the highest accuracy are considered. 
}

\begin{table}[htb]
\centering
\begin{tabular}{| c | c | c | c | c |  }
\hline
	  & \multicolumn{4}{|c|}{$Ht \bar{t}$ }  \\ \cline{2-5}
	  & \multicolumn{2}{|c|}{Higgs} & \multicolumn{2}{|c|}{top}  \\ \cline{2-5}
	  & Enlarge & PELICAN & Enlarge & PELICAN  \\ \hline
10\% & 0.0098 & 0.0082 & 0.0100 & 0.0077  \\ 
20\% & 0.0150 & 0.0132 & 0.0147 & 0.0122  \\ 
30\% & 0.0198 & 0.0189 & 0.0192 & 0.0173  \\ 
40\% & 0.0251 & 0.0272 & 0.0240 & 0.0240  \\ 
50\% & 0.0322 & 0.0424 & 0.0295 & 0.0353  \\ \hline
\end{tabular}
\caption{The values of RMSD for given percentages (from 10\% to 50\%) of most accurate events of the $Ht \bar{t}$ process. The results for Higgs and top obtained from the Enlarged Mask R-CNN method and PELICAN method are shown. \label{tab:rmsdhtt}}
\end{table}

Compared to the results in our earlier works~\cite{Li:2020grn,Guo:2020vvt} where the reconstructed jet momentum is calculated by the vector sum of momenta of all marked particles so that the accuracy is highly affected by the pileup contamination, the method in this work adopts an independent jet branch to predict the ground truth jet momentum. 
It turns out that the jet branch has been trained to implement pileup mitigation in an automatic and efficient way. No further pileup mitigation procedures are required.

\subsection{The detector and pileup effects}
{Although the network is trained on Monte Carlo events without detector effects, it is supposed to also work well on events where detector effects are included. 
For illustration, we apply our network (which is trained on events without detector effects) to events where the energy of final state particles are Gaussian smeared with standard deviation varying from 1\% to 20\% of the total energy. In terms of RMSD as discussed above, the results of the Higgs and top reconstruction accuracy for events with different standard deviations are given in Table~\ref{tag:hemsd} and~\ref{tag:temsd}. 
We can find that the reduction of reconstruction accuracy due to the detector effect is mild, especially for the top.  Given a specific detector configuration, it is also possible for the network to learn the dedicated detector effects (which render the assignment of Higgs/top constituents ambiguous) by training the network on particle-gun MC events (which contain only a single top quark or Higgs in the final state) where the detector effects are included. This can help to mitigate the detector smearing on the momentum precision to some extent. However, possible drawbacks of such a procedure are the method could become detector-dependent and may not be able to learn the feature for the overlapped case as will be discussed later. 
}

\begin{table}[htb]
\centering
\begin{tabular}{| c | c | c | c | c | c | c | c | c | c | c | }
\hline
	  & w/o DS & 0.01$E$ & 0.02$E$ & 0.04$E$  & 0.08$E$ & 0.1$E$ & 0.12$E$ & 0.14$E$ & 0.18$E$ & 0.2$E$ \\ \hline
10\% & 0.0098 & 0.0098 & 0.0101 & 0.0110 & 0.0129 & 0.0133 & 0.0143 & 0.0152 & 0.0167 & 0.0173 \\ 
20\% & 0.0150 & 0.0150 & 0.0153 & 0.0165 & 0.0189 & 0.0197 & 0.0214 & 0.0225 & 0.0250 & 0.0260 \\ 
30\% & 0.0198 & 0.0203 & 0.0202 & 0.0219 & 0.0242 & 0.0258 & 0.0277 & 0.0291 & 0.0325 & 0.0341 \\ 
40\% & 0.0251 & 0.0261 & 0.0259 & 0.0279 & 0.0302 & 0.0322 & 0.0343 & 0.0362 & 0.0404 & 0.0424 \\ 
50\% & 0.0322 & 0.0337 & 0.0333 & 0.0354 & 0.0372 & 0.0400 & 0.0421 & 0.0444 & 0.0496 & 0.0520 \\ 
\hline
\end{tabular}
\caption{The values of RMSD for given percentages (from 10\% to 50\%) of most accurately predicted Higgs in the $Ht \bar{t}$ process. Different columns correspond to the different standard deviations that are taken in the Gaussian smearing of the jet constituent energy $E$. The second column gives the RMSD values without detector smearing (DS). Results are obtained by the enlarged Mask R-CNN method. \label{tag:hemsd}}
\end{table}

\begin{table}[htb]
\centering
\begin{tabular}{| c | c | c | c | c | c | c | c | c | c | c | }
\hline
	  & w/o DS & 0.01$E$ & 0.02$E$ & 0.04$E$  & 0.08$E$ & 0.1$E$ & 0.12$E$ & 0.14$E$ & 0.18$E$ & 0.2$E$ \\ \hline
10\% & 0.0099 & 0.0095 & 0.0100 & 0.0101 & 0.0105 & 0.0101 & 0.0108 & 0.0107 & 0.0112 & 0.0113 \\ 
20\% & 0.0147 & 0.0144 & 0.0149 & 0.0151 & 0.0154 & 0.0152 & 0.0160 & 0.0162 & 0.0167 & 0.0170 \\ 
30\% & 0.0192 & 0.0188 & 0.0194 & 0.0196 & 0.0201 & 0.0200 & 0.0210 & 0.0212 & 0.0219 & 0.0224 \\ 
40\% & 0.0239 & 0.0235 & 0.0243 & 0.0246 & 0.0252 & 0.0251 & 0.0261 & 0.0265 & 0.0275 & 0.0282 \\ 
50\% & 0.0295 & 0.0292 & 0.0302 & 0.0306 & 0.0312 & 0.0310 & 0.0324 & 0.0330 & 0.0342 & 0.0351 \\ 
\hline
\end{tabular}
\caption{Same as Table~\ref{tag:hemsd}, for predicted top in the $Ht \bar{t}$ process. 
\label{tag:temsd}}
\end{table}

{Another important practical issue is the pileup effects during the collision. At the high-luminosity LHC, the average number of pileup interactions per bunch crossing can reach $\langle \mu \rangle  \sim 140-200$. On the other hand, there are pileup mitigation algorithms based on vertex and calorimeter information, which help to suppress the pileup effects in the final data. 
Irrespective of a specific pileup mitigation method, we apply our network (which is trained on events with 50 pileups, denoted by network@PU50) to events with pileup levels varying from 5 to 200 (denoted by PU5 to PU200). 
The results are given in Table~\ref{tab:pileup}. 
The AUC of Mask R-CNN is barely changing for events with pileups smaller than 50. And it decreases steadily with increasing the pileup for $\langle \mu \rangle  \gtrsim 50$.
Moreover, we further fine-tune the network@PU50 on 300 thousand events with 200 pileups and obtain the network@PU200 version of Mask R-CNN. 
In terms of AUC values, the upgraded network has stable performance on events with pileup level up to 200. Meanwhile, the performance is comparable to that of the network@PU50 for low pileup events. 
}

\begin{table}[h!]
\centering
\begin{tabular}{| c | c | c | c | c | c | c | c | c | c | c | c | c | c |}
\hline
    & & PU5 & PU10 & PU20 & PU30  & PU50 & PU80 & PU100 & PU120 & PU150 & PU180 & PU200\\ \hline
\multirow{2}*{network@PU50} & signal Higgs & 0.9724 & 0.9728 & 0.9729 & 0.9732 & 0.9723 & 0.9670 & 0.9589 &0.9446 & 0.9016 &  0.8051 & 0.7037 \\ 
                            & signal top & 0.9743 & 0.9746 & 0.9751 & 0.9756 & 0.9754 & 0.9718 & 0.9659 & 0.9547 & 0.9228 & 0.8732 & 0.8323  \\ \hline
\multirow{2}*{network@PU200} & signal Higgs & 0.9609 & 0.9620 & 0.9636 & 0.9656 & 0.9678 & 0.9691 & 0.9696 & 0.9702 & 0.9705 & 0.9697 & 0.9691 \\ 
                            & signal top & 0.9684 & 0.9690 & 0.9701 & 0.9710 & 0.9723 & 0.9737 & 0.9741& 0.9744 & 0.9744 & 0.9743 &0.9740  \\                        
\hline
\end{tabular}
\caption{AUC values of two versions of enlarged Mask R-CNN being tested on event sample with different pileup levels. 
Besides the version that is used throughout the paper (denoted by network@PU50, because it is trained on events with 50 pileups), the network@PU200 version of Mask R-CNN is obtained by further training the network@PU50 on 300 thousand events with 200 pileups. \label{tab:pileup}}
\end{table}

\subsection{Test on different processes}

\begin{figure}
\centering
\includegraphics[width=0.85\textwidth]{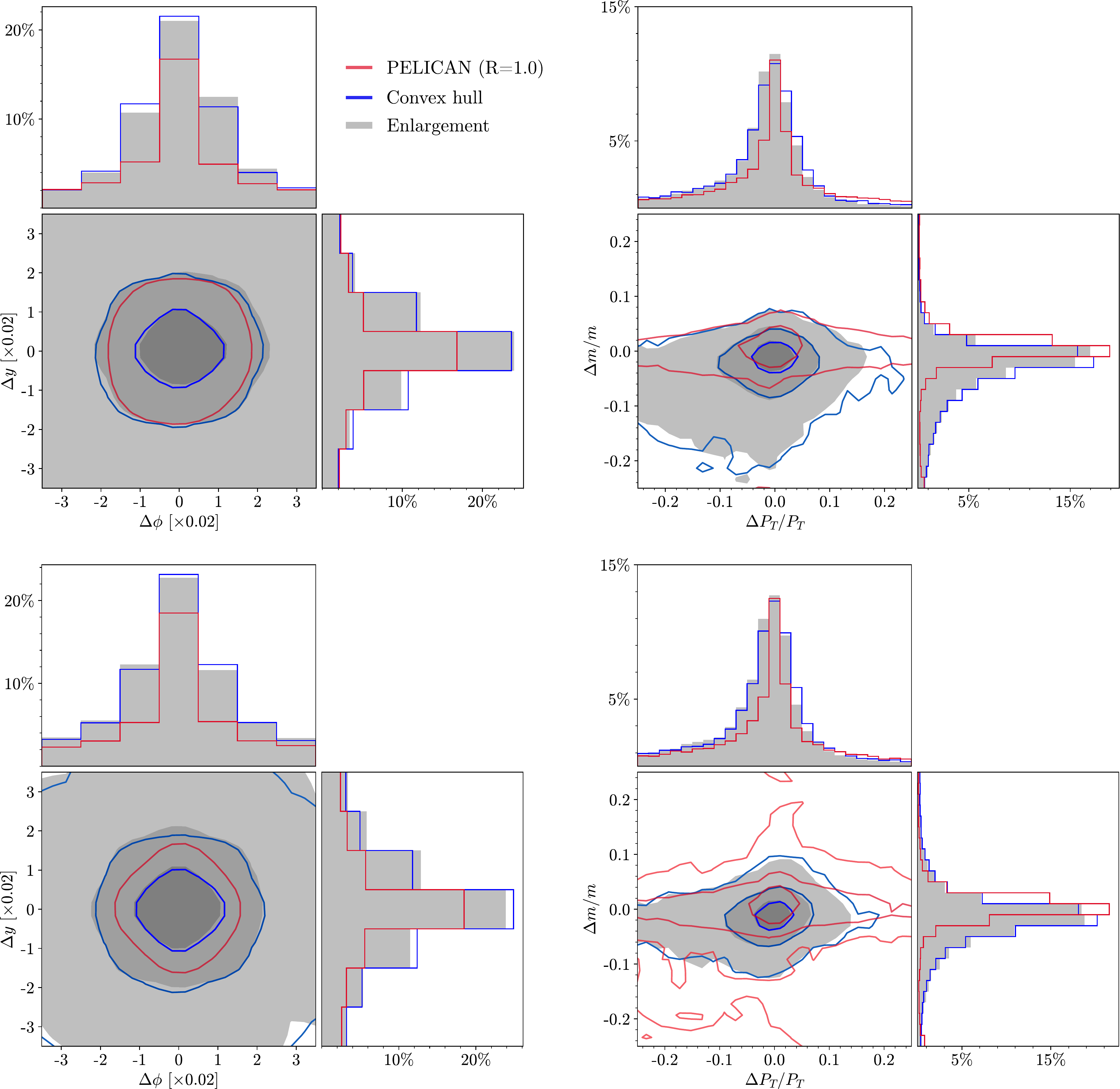}
\caption{\label{fig:hhttx_dist} Similar as figure~\ref{fig:httx_dist} for the reconstructed Higgs (upper panels) and top jets (lower panels) in $H H t \bar{t}$ event sample.
}
\end{figure}  

\begin{figure}
\centering
\includegraphics[width=0.85\textwidth]{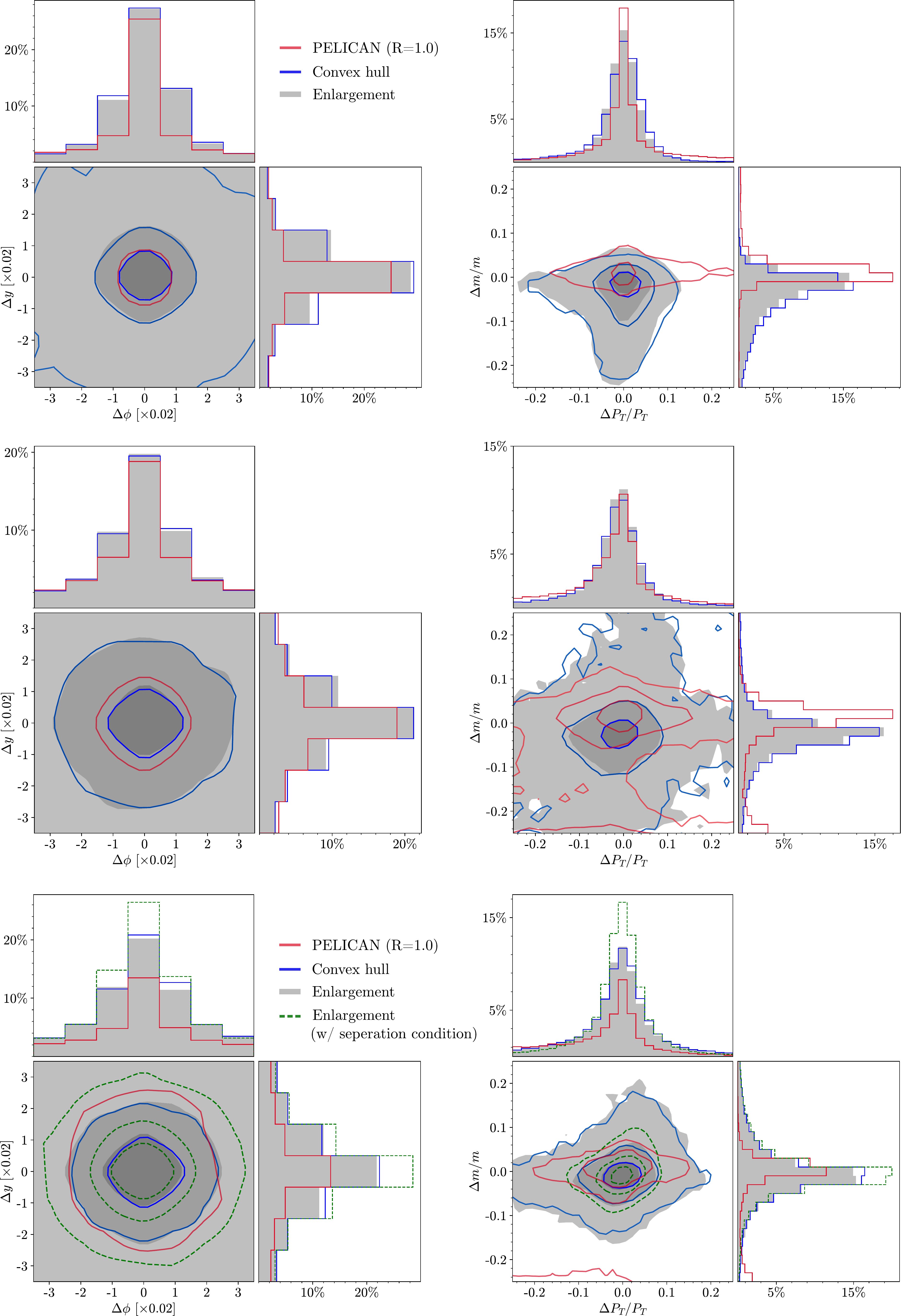}
\caption{\label{fig:others_dist}  Similar as figure~\ref{fig:httx_dist} for reconstructed Higgs jets in the $\tilde{\chi}^{0}_{2} \tilde{\chi}^{0}_{2}$ event sample (upper panels), reconstructed top jets in the $\tilde{t} \bar{\tilde{t}}$ event sample (middle panels), and reconstructed top jets in the $t\bar{t}t\bar{t}$ event sample (lower panels).
}
\end{figure}  

Although the networks have been trained with event samples of $Ht\bar{t}$ process, they can be used as general Higgs and top jets taggers for events of many other processes. 
For demonstration purposes, we showcase their capabilities in other processes at the LHC that produce Higgs and top jets: 1) $p p \to HH t \bar{t}$ in the SM; 2) neutral Higgsino pair production with subsequent decay $\tilde{\chi}^{0}_{2} \to H \tilde{\chi}^{0}_{1}$ in SUSY model; 3) top squark pair production with subsequent decay $\tilde{t} \to t \tilde{\chi}^{0}_{1}$; 4) $ p p \to t \bar{t} t \bar{t}$ in the SM. 
The corresponding accuracy contours are shown in Figure~\ref{fig:hhttx_dist} and Figure~\ref{fig:others_dist}. 
In all cases, we can find that the network performances only slightly depend on the definition of the mask. 
The accuracies for all of the variables ($m,p_{T},y,\phi$) are always slightly worse than those of the $H t\bar{t}$ process. 
Generally speaking, about 40\% of the Higgs/top jets in those test samples can be reconstructed with $\Delta y \sim \Delta \phi \in [-0.04, 0.04]$, $\frac{\Delta m}{m}\in [-0.08,0.05]$ and $\frac{\Delta p_{T}}{p_{T}} \in [-0.1,0.07]$.
The Higgsino pair ($\tilde{\chi}^{0}_{2} \tilde{\chi}^{0}_{2}$) events contain the fewest detectable particles in the final state so that higher accuracies can be obtained for the momentum variables. 
For the stop pair process, only the masses of stop and neutralino are set. There are $\mathcal{O}(10\%)$ of events containing top quark with momentum less than 300 GeV, leading to lower reconstruction efficiencies and decreased accuracies of the momentum variables. 
{The mass predictions for Higgs/top are generally lower for all cases, and the situation is more severe for the stop pair and neutralino pair processes. Because the network tends to predict smaller jet masks, which may drop some jet constituents with non-negligible energy. In the stop pair and neutralino pair processes, such operation happens more frequently, because there are a certain fraction of events with $p_T(H)<200$ GeV and $p_T(t)$<300 GeV.
The quantitative results in terms of RMSD for those processes are provided in Table~\ref{tab:procsrmsd}. 
The fraction of events when Mask RCNN surpasses PELICAN is 18\% for Higgs reconstruction in $HH t\bar{t}$ process, 20\% for top reconstruction in $HH t\bar{t}$ process, 35\% for Higgs reconstruction in $\tilde{\chi}^{0}_{2} \tilde{\chi}^{0}_{2}$ process, 24\% for top reconstruction in $\tilde{t} \bar{\tilde{t}}$ process, and 7\% for top reconstruction in $t \bar{t} t \bar{t}$ process, respectively. 
}

\begin{table}[htb]
\centering
\begin{tabular}{| c | c | c | c | c | c | c |  c | c | c | c |}
\hline
	 & \multicolumn{2}{|c|}{Higgs of $HHt \bar{t}$} & \multicolumn{2}{|c|}{top of $HHt \bar{t}$}  & \multicolumn{2}{|c|}{$\tilde{\chi}^{0}_{2} \tilde{\chi}^{0}_{2}$} & \multicolumn{2}{|c|}{ $\tilde{t} \bar{\tilde{t}}$} & \multicolumn{2}{|c|}{$t\bar{t}t\bar{t}$} \\ \cline{2-11}
	 & Enlarge & PELICAN & Enlarge & PELICAN & Enlarge & PELICAN & Enlarge & PELICAN  & Enlarge & PELICAN  \\ \hline
10\% & 0.0133 & 0.0119 & 0.0127 & 0.0110& 0.0133 & 0.0080 & 0.0184 & 0.0160 & 0.0143 & 0.0152  \\ 
20\% & 0.0210 & 0.0221 & 0.0197 & 0.0198& 0.0207 & 0.0139 & 0.0271 & 0.0261 & 0.0221 & 0.0315  \\ 
30\% & 0.0292 & 0.0416 & 0.0268 & 0.0342& 0.0284 & 0.0236 & 0.0364 & 0.0399 & 0.0306 & 0.0938  \\ 
40\% & 0.0405 & 0.0804 & 0.0359 & 0.0616& 0.0376 & 0.0504 & 0.0489 & 0.0698 & 0.0413 & 0.1712  \\ 
50\% & 0.0593 & 0.1497 & 0.0485 & 0.1108 & 0.0508 & 0.1227 & 0.0731 & 0.1254 & 0.0561 & 0.2438  \\ 
\hline
\end{tabular}
\caption{Same as Table~\ref{tab:rmsdhtt}, for $HHt \bar{t}$, $\tilde{\chi}^{0}_{2} \tilde{\chi}^{0}_{2}$, $\tilde{t} \bar{\tilde{t}}$, and $t\bar{t}t\bar{t}$ processes, respectively. \label{tab:procsrmsd}}
\end{table}

\begin{figure}[h]
\centering
\includegraphics[width=0.8\textwidth]{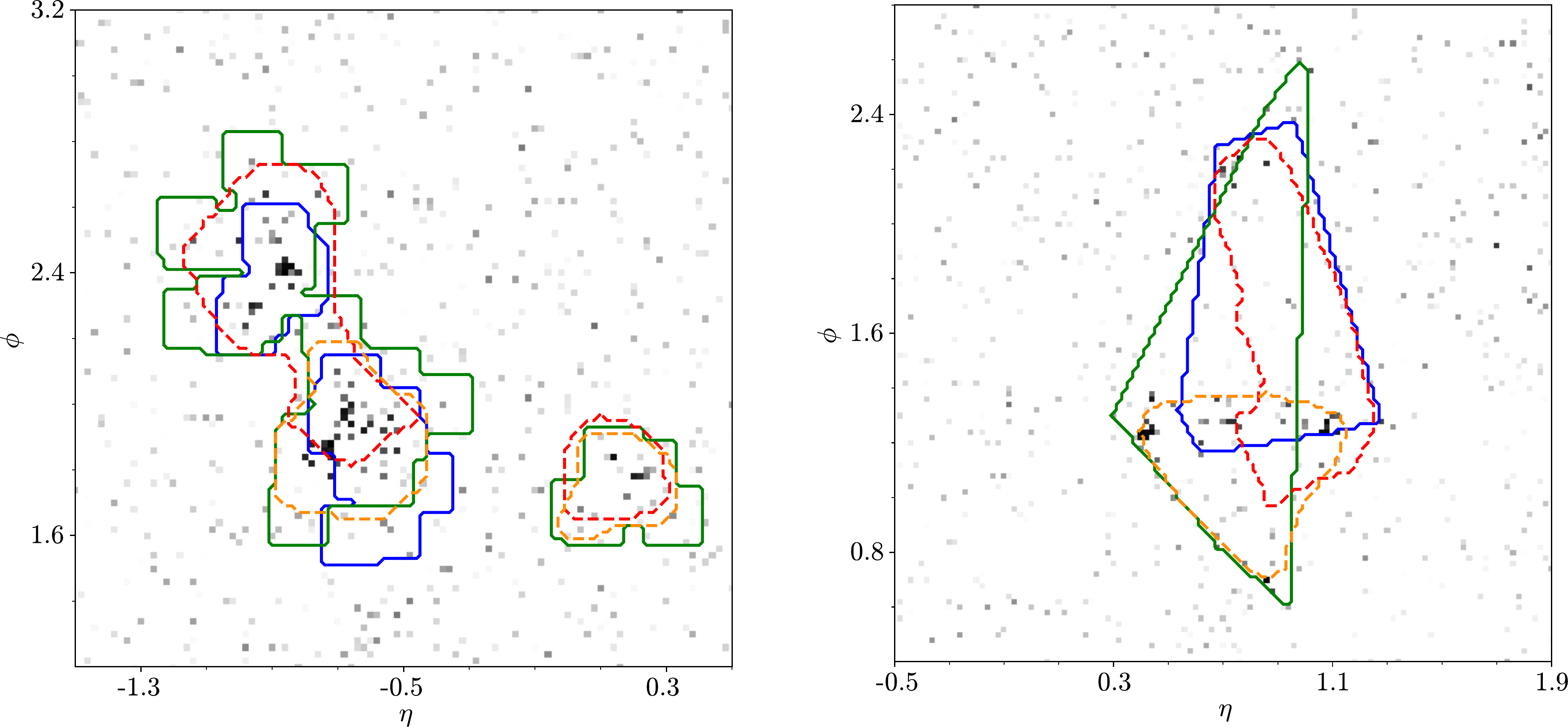}
\caption{\label{fig:overlaps}
Solid lines indicate ground truth masks and dashed lines represent predicted masks.
Overlapping jet areas make the network less accurate.
}
\end{figure}

Since there are multiple jets on an image, 
some of jet areas can be very close to each other
or even overlap with each other.
In such cases,
the sequential recombination algorithms with a large cone size
will end up with merging nearby or overlapping jets into one jet.
In fact, when two final state particles are close to each other,
it is practically impossible to tell if they originate from different ancestors.
Therefore, the overlapping jets will also greatly impede the network accuracy.
To see how much it affects the accuracy,
the network is tested on $t \bar{t}t \bar{t}$ event sample with a separation condition
in which the ground truth coordinates are at least 50 pixels ($\Delta R\simeq$ 1)
away from each other such that a large overlap almost never occurs.
The momenta reconstruction result 
is shown as dashed lines in lower panels of Figure \ref{fig:others_dist} for comparison.
On the other hand, unlike the {PELICAN method which requires the jet clustering},
the network is still mostly capable of finding the correct number of jets.
It seems that the network can identify if there are constituents of multiple jets in an RoI
and outputs the most plausible overlap configuration
based on what it has seen in the training sample,
although the prediction is less accurate as shown in Figure \ref{fig:overlaps} for instance.

{Finally, we apply our network to events of the QCD multijet process $p p \to j j j$ where each jet in the final state is required to have $p_T>200$ GeV.
The Mask R-CNN predicts only 14 Higgs jets and 0 top quark jets out of 10 thousand images
proving its high rejection rate for pure QCD jets.
For comparison, the same event sample is tested by the BDRS Higgs tagging algorithm and the HEPTopTagger top tagging algorithm. With the default setting of jet substructure parameter as given in Refs.~\cite{Butterworth:2008iy,Plehn:2010st}, the BDRS method predicts $\sim$600 Higgs jets with mass in $[$120 GeV,130 GeV$]$, and the HepTopTagger predicts $\sim$550 top jets with mass in $[$150 GeV, 200 GeV$]$, among the 10 thousand event sample. }

\section{Conclusion} \label{conclude}

This work aims to build a deep neural network to label the constituents of target jets among hadronic final state particles based on supervised learning. In particular, the hadronic final state of the top quark can not be identified unambiguously according to the Monte Carlo simulation due to the color interconnection in hadronization. We propose an algorithm based on the ground truth information as well as the {angular distance measure} to determine the top quark jet constituents. 

The Mask R-CNN framework is adopted to detect the Higgs jets and the top quark jets in collision events at the LHC. The network can predict the shape (or mask) of target jets of different kinds on event images. 
The definition of jet shape/mask is not unique, even from a theoretical perspective. 
Two schemes of mask definition are proposed in this work: enlargement mask and convex hull mask. 
More importantly, an additional jet branch is built for predicting the four-momenta of the original partons, in which the pileup mitigation is intrinsically implemented. 

The network is trained on events of the $H t\bar{t}$ process at the LHC, where the transverse momenta of the Higgs and top quarks are required to be $p_{T}(H) > 200$ GeV and $p_{T}(t/\bar{t}) > 300$ GeV. 
Each event is overlaid with an average number of $\langle \mu \rangle =50$ pileup events. 
{Compared with the LorentzNet for jet classification and the PELICAN network for jet momentum regression,
the Mask R-CNN can detect and reconstruct both the Higgs and top jets in a more efficient and accurate way, mainly because it predicts jets with more accurate boundaries. 
}
The networks with two different definitions of the mask have similar performances. 
In terms of two-dimensional distributions on the $\Delta y \times \Delta \phi$ plane and $\frac{\Delta m}{m} \times \frac{\Delta p_{T}}{p_{T}}$ plane, about 60\% of Higgs jets can be reconstructed with $\Delta y \sim \Delta \phi \in [-0.06, 0.06]$, $\frac{\Delta m}{m}\in [-0.08,0.05]$ and $\frac{\Delta p_{T}}{p_{T}} \in [-0.14,0.1]$. And about 60\% of top jets can be reconstructed with $\Delta y \sim \Delta \phi \in [-0.05, 0.05]$, $\frac{\Delta m}{m}\in [-0.07,0.5]$ and $\frac{\Delta p_{T}}{p_{T}} \in [-0.11,0.08]$. 

The generality of the method is demonstrated by applying the Mask R-CNN to processes different from the trained one, including 1) $p p \to HH t \bar{t}$ in the SM; 2) $\tilde{\chi}^{0}_{2} \tilde{\chi}^{0}_{2}$ production with decay $\tilde{\chi}^{0}_{2} \to H \tilde{\chi}^{0}_{1}$ in SUSY model; 3) $\tilde{t} \bar{\tilde{t}}$ production with decay $\tilde{t} \to t \tilde{\chi}^{0}_{1}$; 4) $ p p \to t \bar{t} t \bar{t}$ in the SM. 
In all cases, we find the dependence of the network performance on the mask definition is little. And the network outperforms the PELICAN method in the accuracy of momenta reconstruction, especially for processes with higher visible final state multiplicity. 
In general, the performance is slightly worse than that for the $H t \bar{t}$ process. About 40\% of the Higgs/top jets in those test samples can be reconstructed with $\Delta y \sim \Delta \phi \in [-0.04, 0.04]$, $\frac{\Delta m}{m}\in [-0.08,0.05]$ and $\frac{\Delta p_{T}}{p_{T}} \in [-0.1,0.07]$.

Moreover, we show that the network is capable of detecting the target jets even when they overlap with each other on the event image, although the accuracy of the reconstructed momentum is degraded. 
The network exhibits high background jet rejection power when applied to events of the QCD multijet process. 

Although we have focused on the generalization capability to other processes in this work,
conversely one may have the network to specialize in a particular process through the transfer learning.
By training only the detection head with a small dataset of a certain process,
the accuracy on the process increases while the network becomes rapidly insensitive to other processes.
{The Mask R-CNN method proposed in this work can be used to detect the boosted Higgs/top at the hardware trigger when being loaded to FPGA. Meanwhile, this method can also supplement the conventional analysis, by detecting the Higgs/top and removing the Higgs/top constituent before applying a usual jet clustering algorithm. In the future, we will try to generalize this method to detect all kinds of jets in collider events. Then it can simply replace the jet clustering and indentification algorithms in the conventional data analysis. }

\begin{acknowledgments}
We are grateful to Alexander Bogatskiy for providing the updated version of PELICAN. 
This work was supported by the Natural Science Foundation of Sichuan Province under grant No. 2023NSFSC1329 and the National Natural Science Foundation of China under grant No. 11905149.
S. C. was supported by the Fundamental Research Funds for the Central
Universities, Sichuan University Full-time Postdoctoral Research and Development Fund (No. 2022SCU12118).
\end{acknowledgments}

\appendix
\section{Validation of LorentzNet and PELICAN}
\label{app:val}
To validate the applications of LorentzNet and PELICAN, we try to reproduce the results in Refs.~\cite{Gong:2022lye,Bogatskiy:2022czk}. 
The LorentzNet and the classification component of PELICAN are trained and tested on datasets provided on the website 
\begin{center}
\url{https://osf.io/7u3fk/?view\_only=8c42f1b112ab4a43bcf208012f9db2df}
\end{center} 
and both models were trained by the commands given on their github~\footnote{LorentzNet: \url{https://github.com/sdogsq/LorentzNet-release}}$^,$\footnote{PELICAN: \url{https://github.com/abogatskiy/PELICAN}}. 

\begin{figure}[thb] \centering
\includegraphics[width=0.5\textwidth]{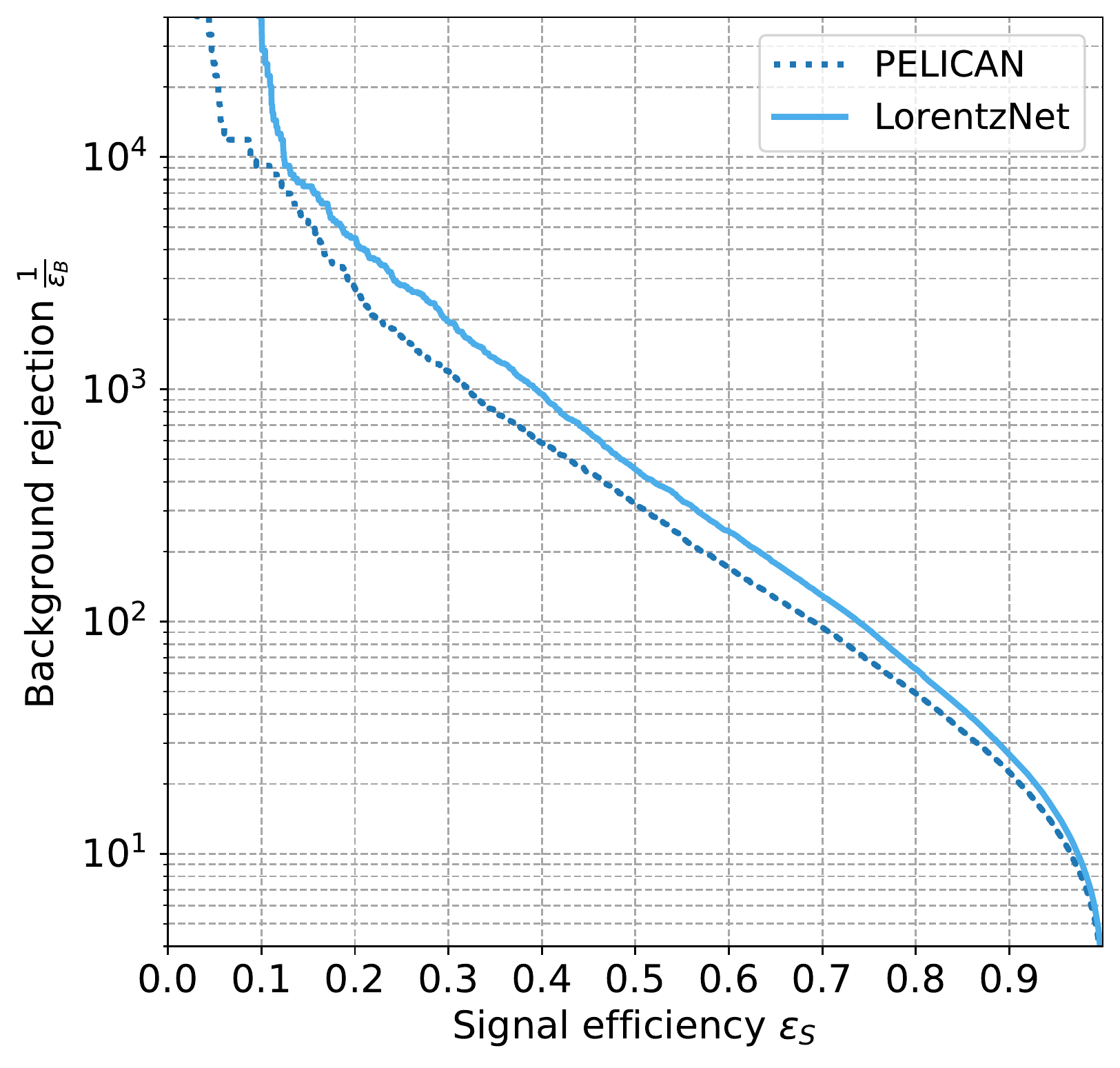}
\caption{ROC curves for top quark jet tagging with LorentzNet and PELICAN. \label{fig:rep_roc}}
\end{figure}

The ROC curves of both methods are shown in Figure~\ref{fig:rep_roc}, for which the signal is the top quark jet and the background is the gluon/light-flavor jets. The [accuracy, AUC] are [0.9417, 0.9865] for LorentzNet and [0.9362, 0.9837] for PELICAN, respectively. The slight degradation of performance for PELICAN is attributed to the limited number of training samples (60000 events are used). 

We also train the PELICAN to predict the four-momentum of W boson for the case without detector effects, as described in Ref.~\cite{Bogatskiy:2022czk}. 
The regression component of PELICAN is trained and tested on the dataset provided on the website
\begin{center}
\url{https://zenodo.org/record/7126443}
\end{center}
The resolutions that we have obtain are: $\sigma_{p_T}=1.12\%$, $\sigma_{m}=1.5\%$, $\sigma_\Psi=0.55$.  Those numbers are close to the values in Ref.~\cite{Bogatskiy:2022czk}. 

\bibliographystyle{jhep}
\bibliography{HTdetection}
\end{document}